\documentclass[superscriptaddress,amsmath,amssymb,aps,preprint,floatfix]{revtex4-2}
\usepackage{graphicx}
\usepackage{subcaption}
\usepackage{float}
\usepackage{dcolumn,multirow}
\usepackage{makecell}
\usepackage{bm}
\usepackage{amsmath,amssymb}
\usepackage{hyperref}
\usepackage{xcolor,soul}
\usepackage[percent]{overpic}

\begin{document}

\title{\textbf{Mapping phase competition in GaAs under pressure and temperature} 
}

\author{Sakshi Patel}
\email{sakshipatelr23@gmail.com}  
\affiliation{Department of Physics, Faculty of Science, The Maharaja Sayajirao University of Baroda, Vadodara-390002, Gujarat, India.}

\author{Paras Patel}
\email{pparas727@gmail.com}  
\affiliation{Department of Physics, Faculty of Science, The Maharaja Sayajirao University of Baroda, Vadodara-390002, Gujarat, India.}

\author{Prafulla K. Jha}
\email{prafullaj@yahoo.com}  
\affiliation{Department of Physics, Faculty of Science, The Maharaja Sayajirao University of Baroda, Vadodara-390002, Gujarat, India.}  

\begin{abstract}
	Pressure-induced transformations in GaAs offer a means to investigate the interplay between crystal structure, electronic behaviour, and phase stability in a technologically important III--V semiconductor.  Here, we have used structure search algorithm combined with first-principles calculations and identified high-pressure competitive phases. The previously proposed $Imm2$ structure lies above the calculated minimum-enthalpy envelope. The static enthalpy calculations with the PBEsol functional predict the transition sequence $F\bar{4}3m \rightarrow Cmcm \rightarrow I\bar{4}m2 \rightarrow C2/m \rightarrow Amm2$, at approximately 12, 14, 34, and 69 GPa, respectively. The initial zinc blende to $Cmcm$ transition is accompanied by a volume reduction of approximately 17.5\% and metallization within the PBEsol description. Phonon calculations support the dynamical stability of all five phases at representative pressures, while elastic calculations characterize their mechanical response. Vibrational free energies are incorporated within the quasiharmonic approximation to examine phase competition over 0--80 GPa and 0--1000 K. These results identify competing high-pressure structures and connect their relative energetics with the electronic, vibrational, and elastic responses of compressed GaAs.
\end{abstract}

\maketitle

High pressure provides a powerful route to access new structural phases and physical phenomena in existing materials \cite{Brazhkin2007,Mao2018}. By modifying interatomic separations and chemical bonding, compression changes the relative stability of competing crystal structures and can stabilize phases that are inaccessible under ambient conditions. These structural changes can substantially alter the electronic, optical, magnetic, and vibrational properties of materials \cite{Mao2018}. Some high-pressure phases can be retained as metastable states after decompression when sufficiently large kinetic barriers inhibit their transformation to the equilibrium structure \cite{Brazhkin2007}. Pressure-induced structural transitions are often accompanied by changes in electronic structure and lattice dynamics \cite{Mujica2003}. In semiconductors, compression can modify band ordering and band gaps, potentially inducing a transition to a metallic or semimetallic state \cite{6}. Pressure also changes phonon frequencies, and the softening of particular modes can signal a dynamical instability toward structural distortion \cite{Kim1999}. A combined analysis of relative enthalpies, electronic structure, and phonon spectra is therefore essential for establishing phase stability and understanding the physical changes associated with pressure-induced structural transitions. 

Binary compounds formed from Group 13 and Group 15 elements constitute the III--V semiconductor family, which includes GaAs, GaP, GaSb, InP, InAs, InSb, AlAs, and AlP. Their diverse band structures and transport properties underpin numerous electronic and optoelectronic applications \cite{Vurgaftman2001}. GaAs is an important member of this family: its direct band gap and high electron mobility make it suitable for light-emitting devices, photovoltaics, and high-speed electronics. At ambient conditions, bulk GaAs adopts the tetrahedrally coordinated zinc-blende (ZB) structure, with space group $F\bar{4}3m$. Compression changes its bonding environment and relative phase energies, providing a means to investigate the relationship between structural transformations, electronic behaviour, and lattice stability \cite{Mujica2003}.

The identification of the high-pressure structures of GaAs has evolved substantially with improvements in diffraction measurements and first-principles calculations. Early electrical-resistance measurements by Minomura and Drickamer reported pressure-induced transitions to a conducting state in 1962 \cite{MINOMURA1962}. Baublitz et al. in 1982, reported first high-pressure phase as orthorhombic symmetry \cite{baublitz1982JAP}, which later assigned to be $Pmm2$ \cite{weir1989}. Zhang and Cohen showed that an orthorhombic structural model was energetically preferred over the rocksalt structure at the  examined compressed volumes \cite{zhang1989cmcm}. Mujica and Needs subsequently demonstrated that this earlier model was unstable against $Cmcm$ and predicted the later structure to become thermodynamically favourable near 12 GPa \cite{mujica1996cmcm}. Angle-dispersive diffraction measurements supported the $Cmcm$ assignment and observed the onset of the ZB to $Cmcm$ transformation at 17.3 GPa and a cinnabar-like phase was also identified during decompression \cite{mcmahon1997cinn}. Electrical transport measurements later reported metallization near 12 GPa, associated with the transformation to the $Cmcm$ phase \cite{Wang2016}. These findings established $Cmcm$ as an important high-pressure phase of GaAs and provided a basis for investigating its subsequent structural evolution.

Despite this progress, the structural evolution of GaAs beyond $Cmcm$ remains less consistently described. In particular, the proposed orthorhombic $Imm2$ phase has been a subject of continued debate in terms of experimental and theoretical supports. Weir et al. assigned a second high-pressure phase to $Imm2$ above 24 GPa \cite{weir1989}. However, subsequent angle-dispersive diffraction measurement by McMahon and Nelmes did not support a separate transition in this pressure region and instead identified the previously assigned GaAs-II and GaAs-III phases as a single $Cmcm$ phase \cite{mcmahon1996new}. Nevertheless, $Imm2$ continued to appear in theoretical investigations. 
Durandurdu and Drabold obtained a $Cmcm$ to $Imm2$ transition near 57 GPa in constant pressure relaxation simulations, while acknowledging that proposed $Imm2$ phase had not been observed in the subsequent diffraction experiments \cite{durandurdu2002}. Subsequently, Pluengphon et al. predicted $Imm2$ between pressure range of 37 -- 88 GPa \cite{Pluengphon2014}. Thus, although theoretical studies have identified $Imm2$ as a candidate high-pressure phase, its equilibrium stability in GaAs remains uncertain.

This uncertainty surrounding $Imm2$ is not restricted to GaAs only. For other III-V compounds, this structure has been theoretically predicted for GaP, InP, and InAs by Mujica and Needs \cite{theoretical_Imm2_IIIV}, for InSb by Guo et al. \cite{InSb_Imm2}, and for GaSb by Belarbi et al. \cite{GaSb_Imm2}. Furthermore, Christensen et al. showed that the predicted high-pressure structural ordering can be sensitive to the inclusion of generalized-gradient corrections when competing structures have small energy differences \cite{solidsunderpressure}. These considerations reinforce the need to compare competing structures directly rather than assuming that an $Imm2$ phase necessarily forms part of the equilibrium sequence of GaAs.

Consequently, the structural evolution of GaAs beyond the well-established \(Cmcm\) phase remains an important open question. Conflicting experimental assignments, differing theoretical predictions, and the sensitivity of the proposed \(Imm2\) stability to computational approximations highlight the need for a systematic reassessment of competing high-pressure structures. In this work, we combine first-principles calculations within density functional theory (DFT) and crystal structure prediction using CALYPSO \cite{PSO,CALYPSO} to investigate the high-pressure structural evolution of GaAs. Our calculations do not identify \(Imm2\) as a thermodynamically competitive phase within the investigated pressure range. Instead, we identify three alternative structures, \(I\bar{4}m2\), \(C2/m\), and \(Amm2\), that exhibit lower enthalpies within their respective stability ranges. We further investigate their dynamical and mechanical stability and electronic properties. Finally, employing the quasiharmonic approximation (QHA), we construct the pressure–temperature phase diagram of GaAs and elucidate the temperature dependence of its high-pressure phase boundaries. These calculations employ the PBEsol exchange-correlation functional and the projector augmented-wave (PAW) method \cite{PAW}, as implemented in VASP \cite{VASP}. Further computational details, including structural optimization, phonon calculations, and the QHA methodology, are provided in the Supplemental Material (SM) \cite{SM}. Our results provide an alternative description of the high-pressure phase evolution of GaAs and a comprehensive assessment of its structural, dynamical, mechanical, and electronic properties.

\begin{figure}[h]
	\centering
	
	\begin{minipage}{1\columnwidth} 
		\raggedright 
		{\fontsize{12}{14}\selectfont} \ 
		\includegraphics[width=\linewidth]{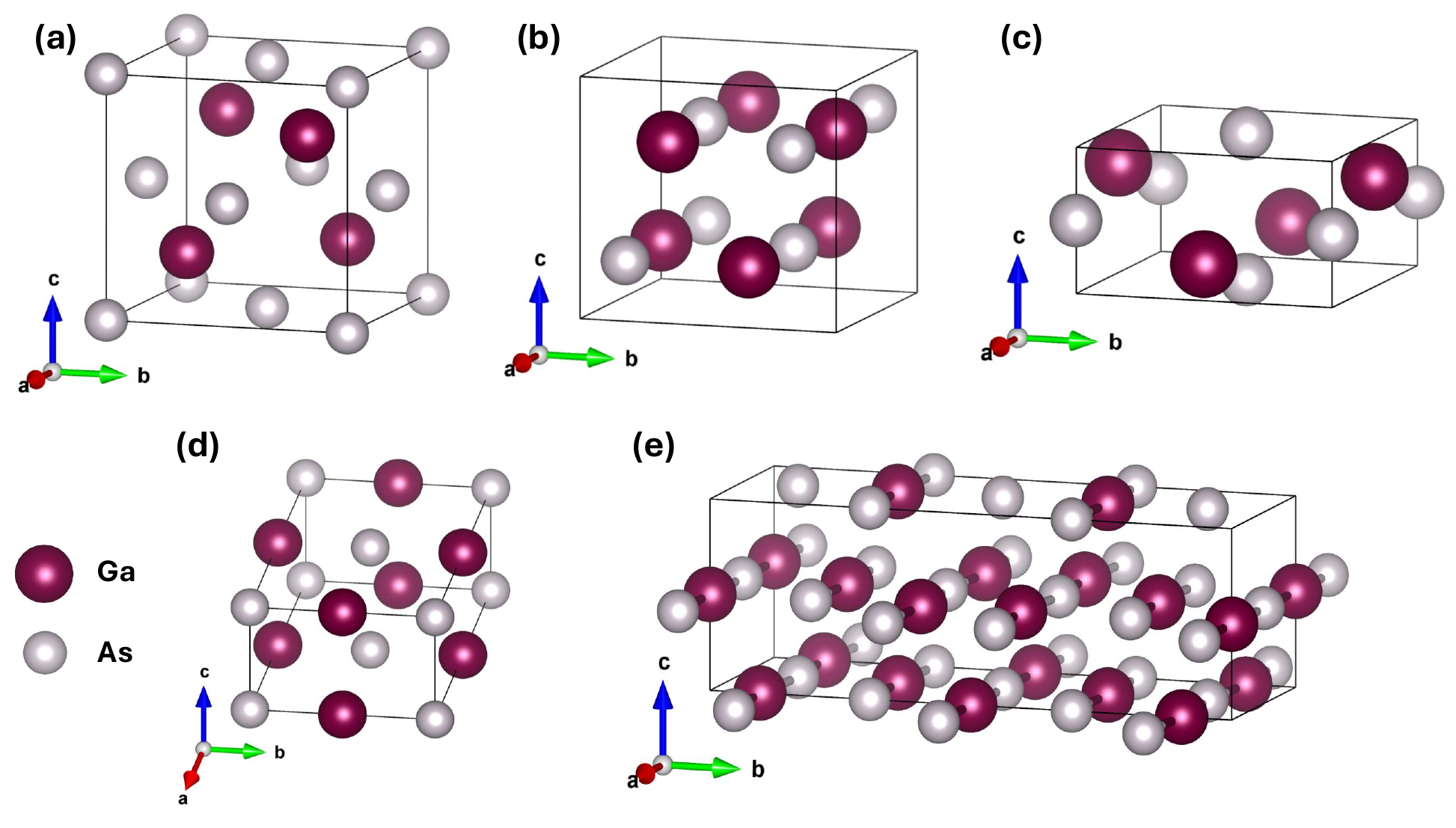} 
	\end{minipage}
	\caption{The structural view of conventional unit cell of (a) Cubic $F\bar{4}3m$ at 0 GPa (b) Orthorhombic $Cmcm$ at 12 GPa (c) Tetragonal $I\bar{4}m2$ at 14 GPa (d) Monoclinic $C2/m$ at 35 GPa and (e) Orthorhombic $Amm2$ at 70 GPa.}
	
	\label{fig:str}
\end{figure}

\begin{figure}[h]
	\centering
	
	\begin{minipage}{0.30\textwidth}
		\makebox[\linewidth][l]{\textbf{(a)}}\\[2mm]
		\includegraphics[width=\linewidth]{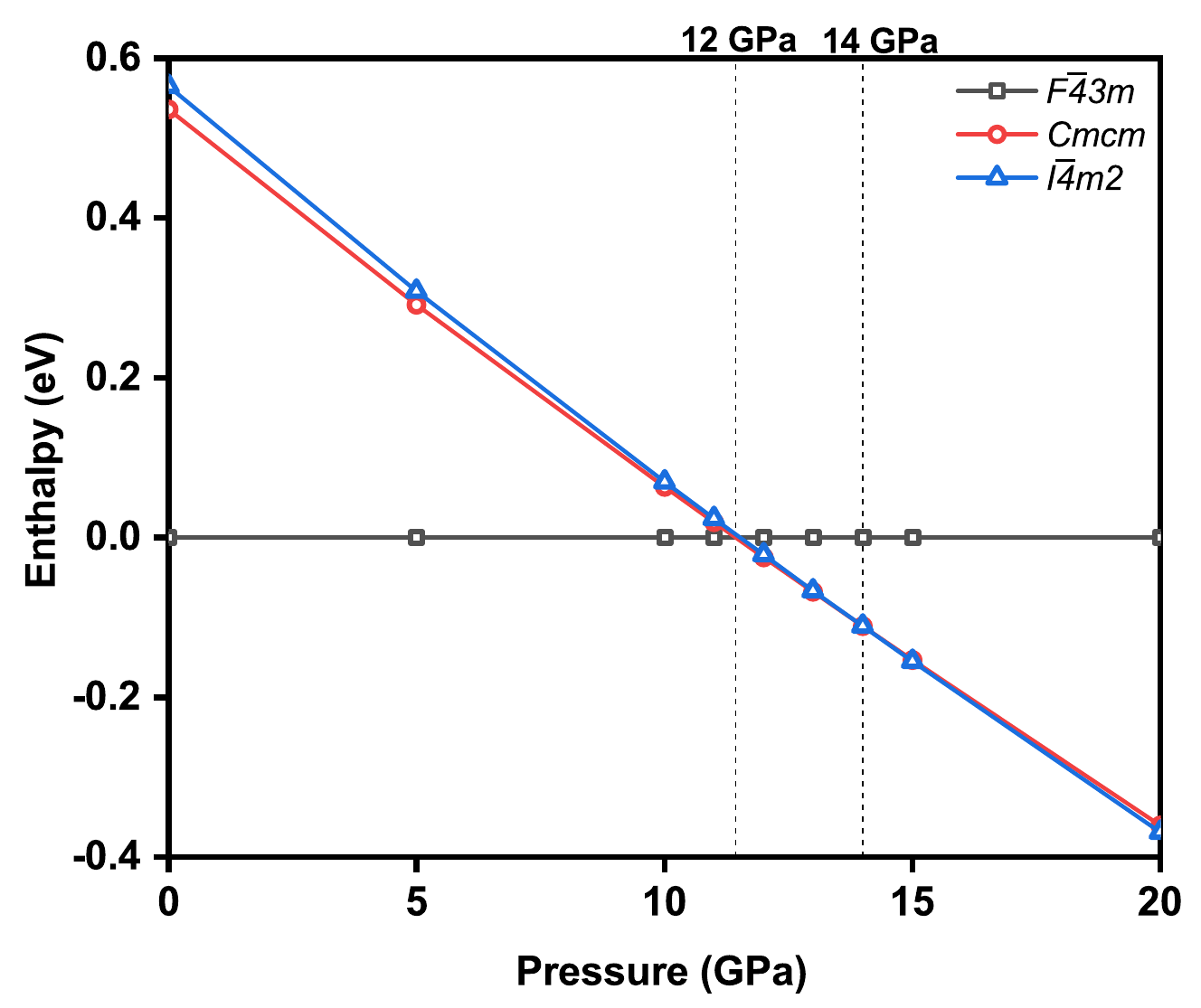}
	\end{minipage}
	\hfill
	\begin{minipage}{0.30\textwidth}
		\makebox[\linewidth][l]{\textbf{(}\textbf{b}\textbf{)}}\\[2mm]
		\includegraphics[width=\linewidth]{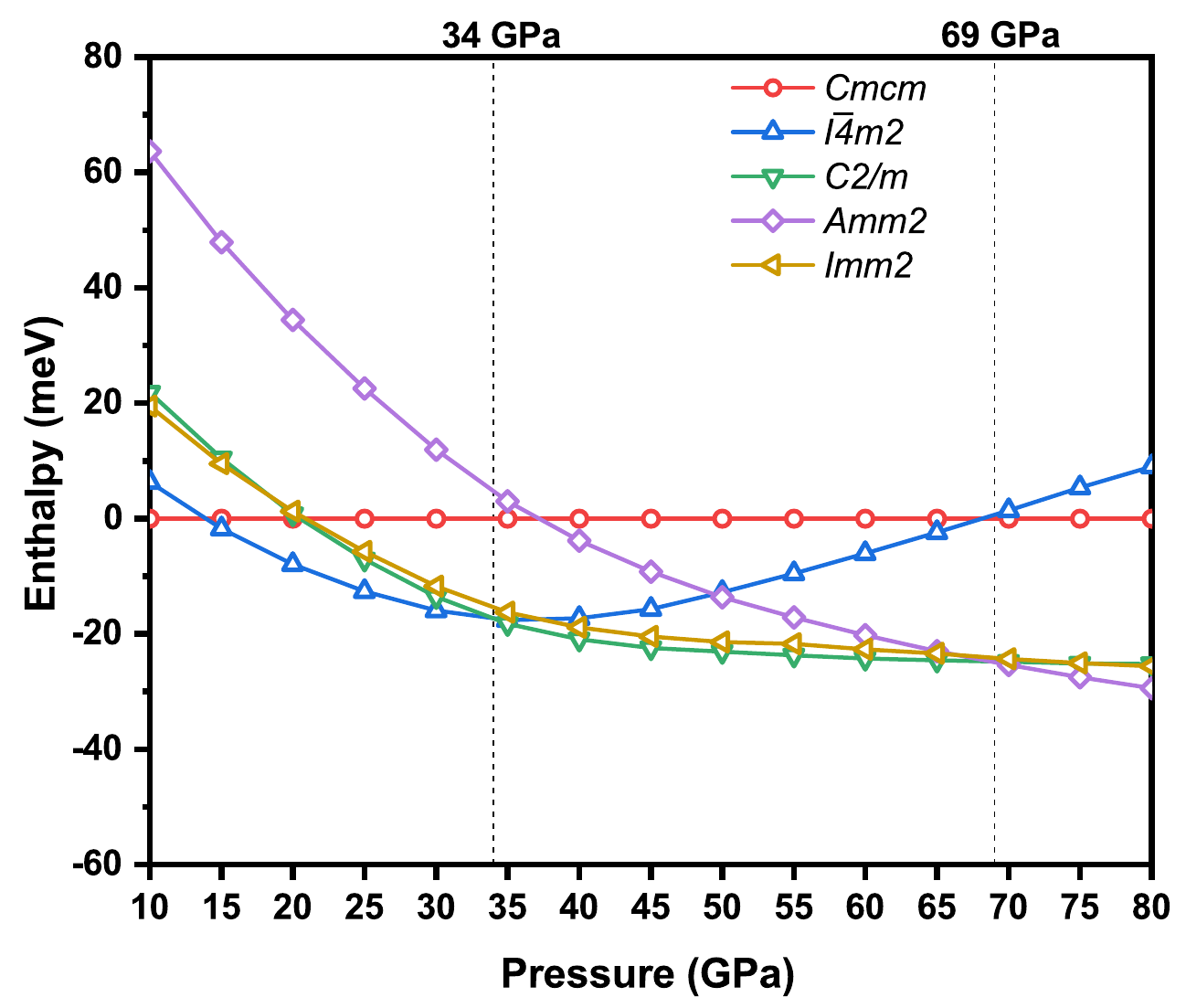}
	\end{minipage}
	\hfill
	\begin{minipage}{0.30\textwidth}
		\makebox[\linewidth][l]{\textbf{(}\textbf{c}\textbf{)}}\\[2mm]
		\includegraphics[width=\linewidth]{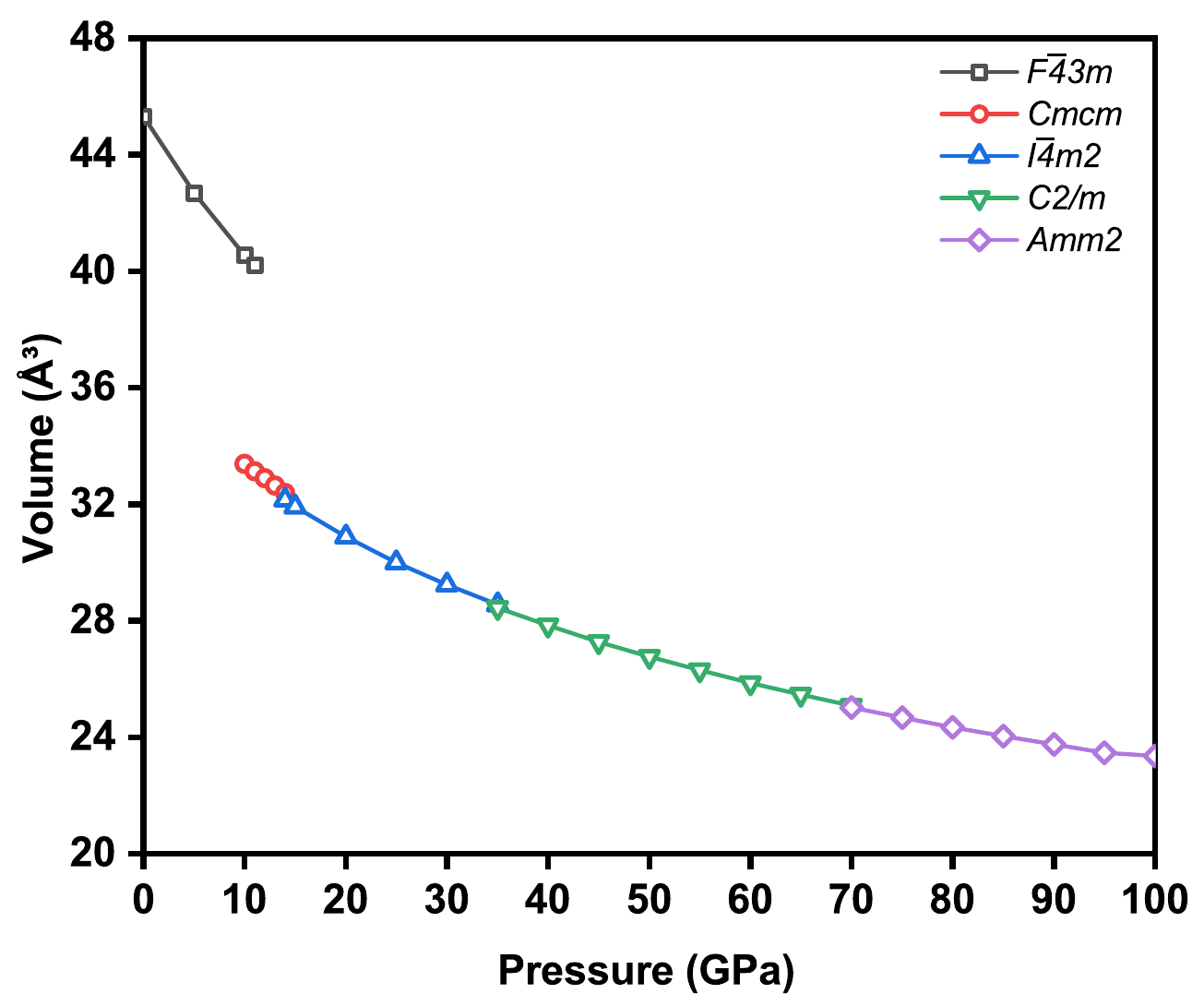}
	\end{minipage}
	
	\caption{(a) Relative enthalpy per formula unit of GaAs low pressure phases referenced to $F\bar{4}3m$ (b) Relative enthalpy per formula unit of GaAs phases at higher pressure referenced to $Cmcm$ and (c) Variation of volume per formula unit of GaAs with pressure for predicted stable phases.}
	\label{fig:enthalpy}
\end{figure}

At ambient condition (zero pressure),  GaAs adopts the cubic ZB structure with space group $F\bar{4}3m$. In the conventional cubic cell, Ga and As atoms occupy Wyckoff positions $4a$ and $4c$ with representative fractional coordinates $(0,0,0)$ and $(\tfrac14,\tfrac14,\tfrac14)$, respectively. The structure consists of two interpenetrating FCC sublattices, with each Ga atom tetrahedrally coordinated by four nearest-neighbour As atoms and vice versa, as illustrated in Fig.~\ref{fig:str}. The relaxed lattice parameter at 0 GPa is $5.65~\mathrm{\AA}$, in good agreement with reported experimental values \cite{lattice_value}\cite{Vurgaftman2001} and with the previous theoretical GGA-PBE result \cite{lattice_ZB_Theory}. The present PBEsol value is slightly closer to the experimental value, confirming the accuracy of the calculated equilibrium structural parameters. The detailed structural parameters are provided in the Table~S1 of SM \cite{SM}. The structural view Fig.~\ref{fig:enthalpy} (a and b) compares the pressure-dependent enthalpies of the investigated phases. 

The static enthalpy curves predict the pressure-induced sequence $F\bar{4}3m \rightarrow Cmcm \rightarrow I\bar{4}m2 \rightarrow C2/m \rightarrow Amm2$ among the investigated structures. The cubic ZB phase has the lowest enthalpy at zero pressure and under low compression. Near 12 GPa, the $Cmcm$ phase becomes energetically favoured, followed by a transition to the tetragonal $I\bar{4}m2$ phase near 14 GPa. These closely spaced transition pressures indicate a narrow pressure interval in which $Cmcm$ has the lowest static enthalpy. The $I\bar{4}m2$ phase remains energetically favourable upto approximately 34 GPa, above which transition occurs to the $C2/m$ phase. The transition from the tetragonal structure to the monoclinic structure represents further change in crystallographic symmetry and lattice geometry. On further increasing pressure, the $C2/m$ phase transit to an orthorhombic $Amm2$ phase at 69 GPa, which retains the lowest enthalpy up to the maximum investigated pressure of 80 GPa.

The calculated PBEsol crossing between the ZB and $Cmcm$ phases occurs near 12 GPa. This value falls within the thermodynamic transition region of $12\pm1.5$ GPa inferred by Besson et al. at 300 K \cite{besson1991high} and is numerically close to the transition reported at 12.2 GPa from Raman and electrical conductivity measurements under hydrostatic conditions \cite{2021_raman}. Higher transformation onsets of 16.6 and 17.3 GPa have been observed in diffraction experiments \cite{weir1989,ono2018_xray}. The exchange-correlation approximation also affects the calculated boundary: Shahi et al. obtained transition pressures of 12.4, 14.1, and 17.1 GPa using LDA, PBE, and SCAN, respectively \cite{SCAN_PT}.

The predicted structural sequence beyond $Cmcm$ also differs from earlier proposals. Although energy-dispersive diffraction originally suggested an $Imm2$ phase above approximately 24 GPa \cite{weir1989}, subsequent angle-dispersive measurements reinterpreted the relevant diffraction patterns using $Cmcm$ and found no evidence for a distinct transition at that pressure \cite{mcmahon1997cinn}. Nevertheless, $Imm2$ remained a candidate in subsequent first-principles calculations \cite{durandurdu2002}. Recent work on yttrium has also demonstrated the improved accuracy of the $r^2$SCAN functional over PBE in predicting pressure-induced phase transitions, with transition pressures in close agreement with experimental observations \cite{Parassir_Y}. Our calculations instead favour $C2/m$ in the corresponding high-pressure interval. Independent relaxation of $C2/m$ and $Imm2$ using $r^2$SCAN preserves this ordering at 35 GPa, with $C2/m$ lower in enthalpy by approximately 5 meV per GaAs formula unit (See Fig. S1 in SM \cite{SM}). This comparison supports the identification of $C2/m$ as a competing high-pressure structure within the candidate set examined.

The pressure dependent equilibrium volumes provide further insight into the structural evolution of GaAs. As shown in Fig.~\ref{fig:enthalpy}(c), the volume decreases smoothly with increasing pressure within each phase. The largest discontinuity occurs at the predicted $F\bar{4}3m$ to $Cmcm$ transition near 12 GPa, where the calculated volume decreases by approximately 17.5\%. This magnitude closely matches the 17.2\% collapse reported experimentally for the ZB to orthorhombic transformation \cite{weir1989}. The structural transformation produces a substantial reorganization of the bonding network. At 12 GPa, ZB has four equivalent Ga--As neighbors at approximately 2.348 \AA, whereas $Cmcm$ has six Ga--As neighbors distributed among four distance groups between 2.535 and 2.756 \AA. Although the individual Ga--As contacts become longer, the increased coordination permits more efficient packing, which leads to a considerably denser structure. By comparison, no similar pronounced volume discontinuities are evident at the subsequent transitions near 14, 34, and 69 GPa. These later transformations therefore involve changes in crystallographic symmetry accompanied by comparatively modest changes in equilibrium volume.

The phonon dispersion curves presented in Fig. S3 of the SM \cite{SM} show no imaginary phonon frequencies along the high-symmetry Brillouin zone paths for any of the predicted equilibrium phases over their respective investigated pressure ranges, confirming their dynamical stability. In particular, the phonon results for $Cmcm$, $I\bar{4}m2$, $C2/m$, and $Amm2$ complement the enthalpy analysis, providing vibrational evidence that the energetically favoured high-pressure structures are locally stable within the harmonic approximation. We have also calculated phonon frequency for the earlier proposed $Imm2$ phase at 20 GPa to establish its stability, however we observe imaginary frequency at the center of the BZ (See Fig. S2 in SM \cite{SM}). 
The zone-centre optical phonons of the $I\bar{4}m2$, $C2/m$, and $Amm2$ phases were further classified by their irreducible representations to determine the Raman and infrared (IR) selection rules. For $I\bar{4}m2$, the optical-mode representation is $\Gamma_{\mathrm{opt}}=B_2+E$, where both symmetry species are Raman and IR active, and the $E$ mode is doubly degenerate. For $Amm2$, the optical modes decompose as $\Gamma_{\mathrm{opt}}=9A_1+4A_2+5B_1+9B_2$. All 27 optical modes are Raman active, whereas the IR-active representation is $9A_1+5B_1+9B_2$; the four $A_2$ modes are IR inactive. For the calculated $C2/m$ structure, the optical-mode representation is $\Gamma_{\mathrm{opt}}=A_u+2B_u$. These three modes are IR active, and no first order Raman active optical modes are predicted for this structure. The calculated frequencies and symmetry assignments are listed in Table~S2 of the SM \cite{SM}. These distinct selection rules provide potential spectroscopic signatures for distinguishing the competing high-pressure phases. 

Complementing the phonon analysis, mechanical stability was assessed from the elastic response to infinitesimal homogeneous strains. The calculated elastic coefficients satisfy the corresponding Born stability criteria for all five phases at the investigated pressures, supporting their mechanical stability \cite{Mouhat2014,elastic_monoclinic}. The stability criteria for each crystal system and the expressions used to derive the elastic properties, along with the calculated independent elastic stiffness coefficients $C_{ij}$, the Voigt--Reuss--Hill averaged bulk modulus $B$ and shear modulus $G$, and the corresponding isotropic Young's modulus $E$ and Poisson's ratio $\nu$, for all five phases, are provided in Table~S3 of the SM \cite{SM}. For the ZB phase at 0 GPa, the calculated elastic constants show good agreement with both the experimental results \cite{elastic_ZB_Exp} and previous theoretical calculations using the GGA-PW91 functional \cite{elastic_ZB_Theory}. 

Further to examine how lattice vibrations affect the static enthalpy sequence, we evaluated the Gibbs free energies of the competing GaAs phases within the quasiharmonic approximation (QHA) \cite{Madhavimam_QHA}, as implemented in Phonopy \cite{Phonopy}. QHA accounts for vibrational zero-point and thermal contributions while incorporating the volume dependence of harmonic phonon frequencies \cite{mounet2005first}. For each phase, the equilibrium Gibbs free energy at pressure $P$ and temperature $T$ is obtained by minimizing \cite{gibbs}
\[
G(P,T)=\min_V\left[F(V,T)+PV\right],
\]
where $F(V,T)$ is the Helmholtz free energy. For comparison between structures, the free energies and volumes are expressed per GaAs formula unit.

\begin{figure}[h]
	\centering
	
	\begin{minipage}{1\columnwidth} 
		\raggedright 
		{\fontsize{12}{14}\selectfont} \ 
		\includegraphics[width=\linewidth]{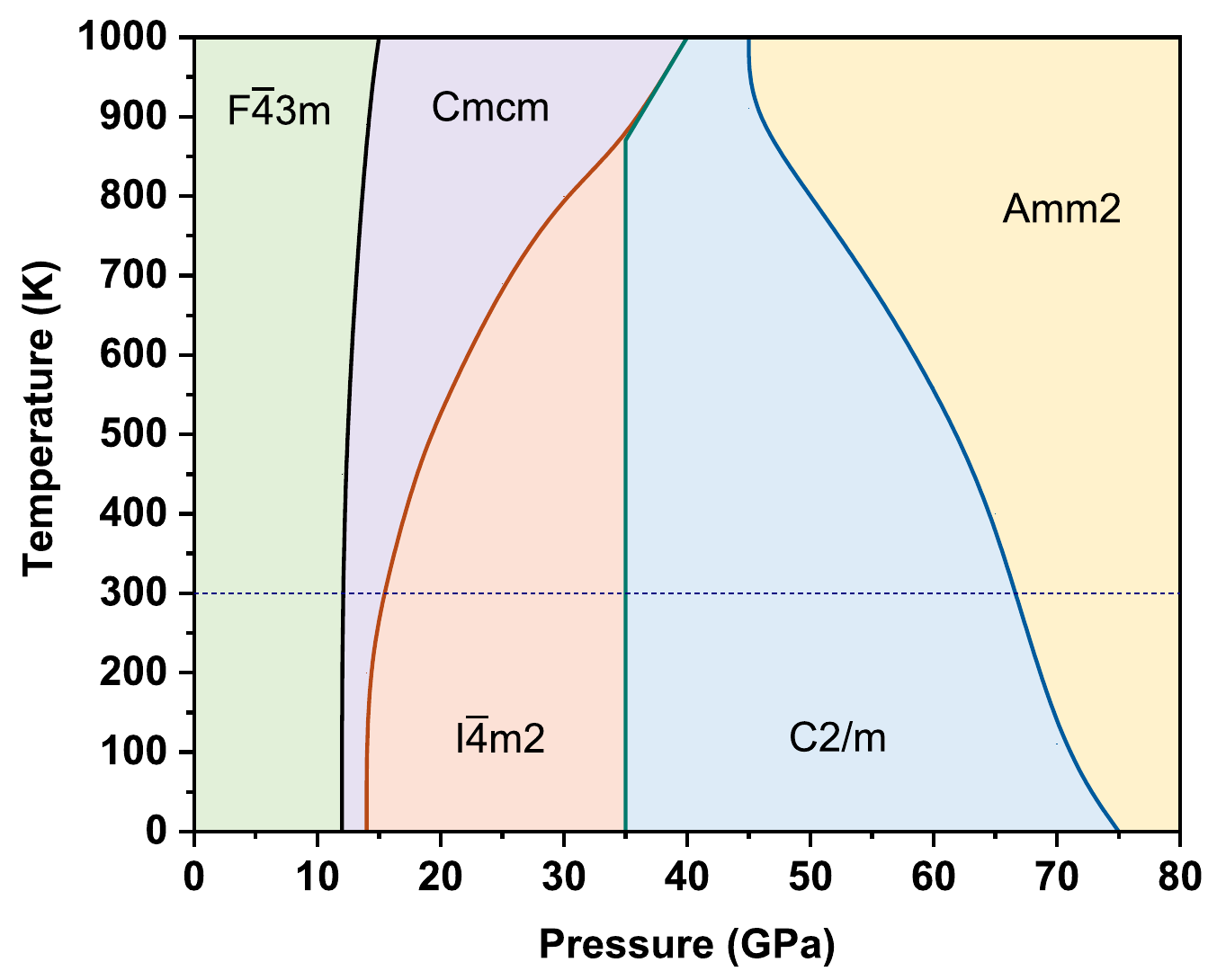} 
	\end{minipage}
	\caption{Pressure--temperature phase diagram of GaAs constructed using the QHA. The coloured regions denote the predicted thermodynamically stable phases, and the solid curves represent the phase boundaries determined from the Gibbs free energy. The horizontal dashed line indicates room temperature (300 K).}
	
	\label{fig:qhaPT}
\end{figure}

At fixed pressure and temperature, the thermodynamically preferred phase has the lowest Gibbs free energy among the considered candidates. A boundary between phases $\alpha$ and $\beta$ occurs where
\[
G_\alpha(P,T)=G_\beta(P,T),
\]
provided that no other considered phase has a lower Gibbs free energy. Comparing these free energies allows us to assess whether vibrational contributions shift the transition pressures or modify the narrow stability interval predicted for $Cmcm$ by the static enthalpy calculations. As shown in Fig.~\ref{fig:qhaPT}, the phase boundaries exhibit distinct temperature-dependent trends. With increasing temperature, the ZB--$Cmcm$ phase boundary shifts slightly toward higher pressure. Similarly, the $Cmcm$--$I\bar{4}m2$ transition pressure increases from approximately 14 GPa at 0 K to about 35 GPa at $\sim$870 K. In contrast, the $I\bar{4}m2$--$C2/m$ phase boundary remains nearly unaffected within the QHA framework. The $C2/m$--$Amm2$ phase boundary exhibits the opposite trend, with the transition pressure decreases as the temperature increases.

To further characterize the predicted structures, simulated X-ray diffraction patterns were obtained for all five phases. The calculated patterns exhibit distinct diffraction features, particularly in terms of peak positions and splitting, which help distinguish the different crystal structures. Comparisons with available experimental patterns for the ZB and $Cmcm$ phases are also provided, showing qualitative correspondence with the reported diffraction features (see SM \cite{SM}).

\begin{figure}[h]
	\centering
	
	\begin{minipage}{1\columnwidth} 
		\raggedright 
		{\fontsize{12}{14}\selectfont} \ 
		\includegraphics[width=\linewidth]{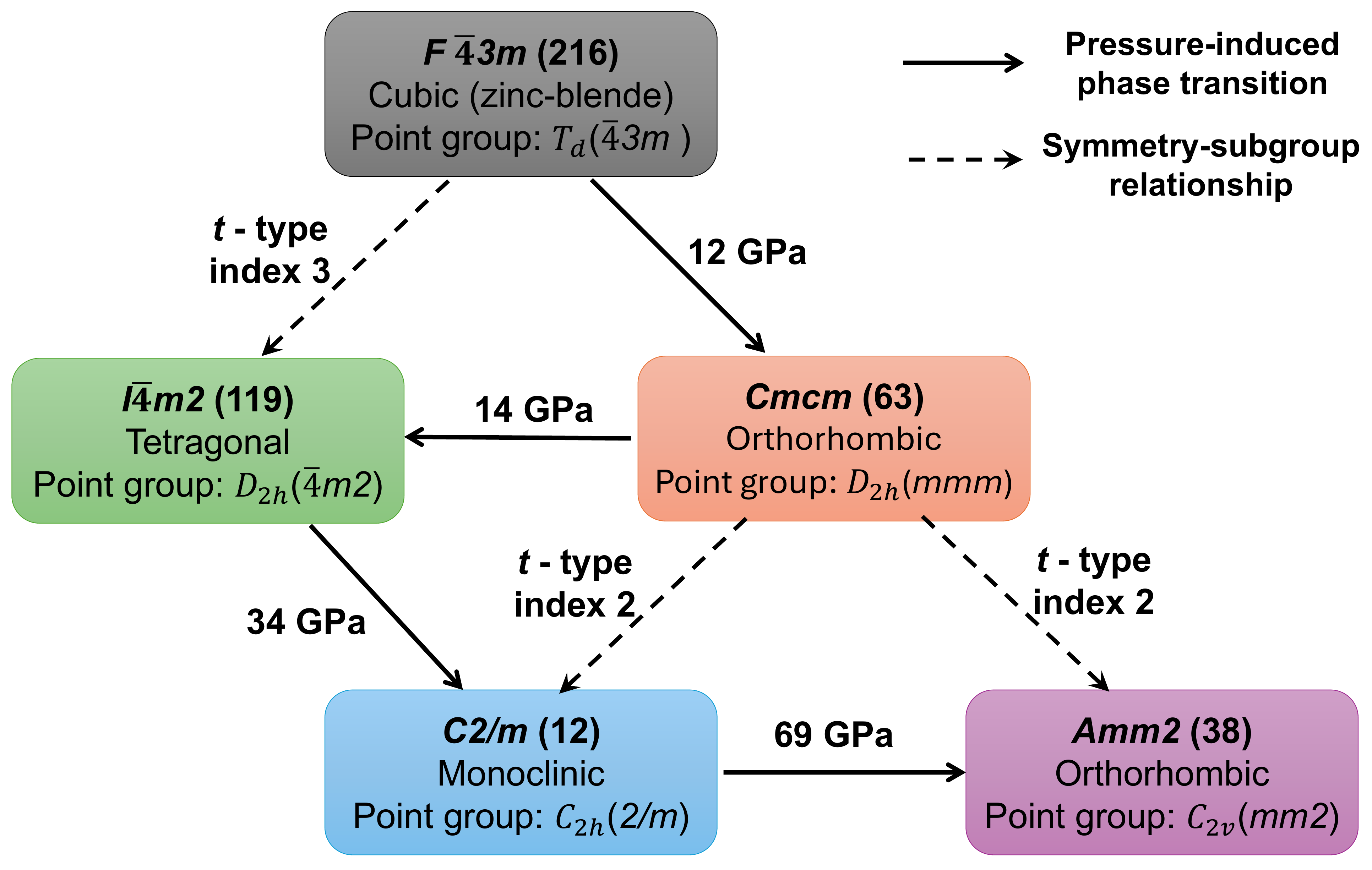} 
	\end{minipage}
	\caption{Schematic representation of the pressure-induced phase transition sequence in GaAs and the group-subgroup relationships among selected structures.}
	
	\label{fig:sym}
\end{figure}

To examine the symmetry relationships underlying this phase transition sequence, the space groups of the predicted phases were analyzed in terms of their possible group-subgroup relationships. Such relations can help identify symmetry elements lost or retained during a structural transformation and provide a crystallographic framework for understanding how the symmetry of the structures may change during compression. The crystallographic relationships among the predicted GaAs phases were examined using the Bilbao Crystallographic Server (BCS) \cite{BCS}. The analysis identifies possible translationengleiche ($t$-type) subgroup relationships between selected structures, corresponding to symmetry reduction while preserving the translational symmetry of the parent group. However, the thermodynamic phase sequence obtained from the enthalpy calculations does not constitute a continuous group-subgroup chain (see Fig. \ref{fig:sym}). Although a $t$-type subgroup relationship of index 3 connects the cubic ($F\bar{4}3m$) and tetragonal ($I\bar{4}m2$) space groups, indicating a reduction in crystallographic symmetry while preserving the translation subgroup, the calculated enthalpy sequence proceeds through the intermediate orthorhombic $Cmcm$ phase. Similarly, the $Cmcm$ space group is related to the monoclinic $C2/m$ and orthorhombic $Amm2$ phases through possible $t$-type subgroup relationships. These relationships indicate symmetry reduction with an index of 2 for each transformation. However, the calculated enthalpy sequence proceeds differently, rather than following these subgroup relationships as a single continuous chain. Therefore, the identified group-subgroup relations provide a complementary crystallographic description of the predicted phase transformations but do not, establish the actual thermodynamic transition pathway. 

The electronic structures of the five GaAs phases were examined using primitive-cell band structures along the corresponding high-symmetry paths and densities of states (DOS) at representative pressures within their energetically favourable regions. At 0 GPa, the ZB  phase exhibits a direct band gap, with both the valence-band maximum and conduction-band minimum located at $\Gamma$-point (See Fig.~S4 of SM \cite{SM}). The calculated band gap is approximately 0.38 eV. Although the direct-gap character is reproduced, its magnitude is substantially smaller than the experimental low-temperature gap of approximately 1.52 eV \cite{Vurgaftman2001}. In contrast to the semiconducting ZB phase, the calculated $Cmcm$, $I\bar{4}m2$, $C2/m$, and $Amm2$ structures exhibit gapless electronic spectra at the investigated pressures. Their band structures contain states crossing the Fermi level. These features indicate metallic electronic character within the present PBEsol description and associate the predicted structural transformations with the loss of the semiconducting gap.

The electronic and bonding properties of GaAs were examined using LOBSTER \cite{LOBSTER}. In the $F\bar{4}3m$ phase sampled Kohn-Sham gap decreases moderately from 1.13 eV at 10 GPa to 1.07 eV at 13 GPa, while the DOS at the Fermi level remains zero. In contrast, $Cmcm$ exhibits partially occupied electronic states and a finite Fermi-level DOS at every investigated pressure. Thus, metallization near 12 GPa is associated with the selection of the metallic Cmcm structure due to energetic favour, rather than complete gap closure within $F\bar{4}3m$. The metallic states have predominantly As 4p character, with a secondary Ga 4p contribution (See Fig. S6 of SM \cite{SM}). 

\begin{figure*}[t]
	\centering
	
	\begin{minipage}{0.30\textwidth}
		\makebox[\linewidth][l]{\textbf{(a)}}\\[1mm]
		\centering
		\includegraphics[width=\linewidth]{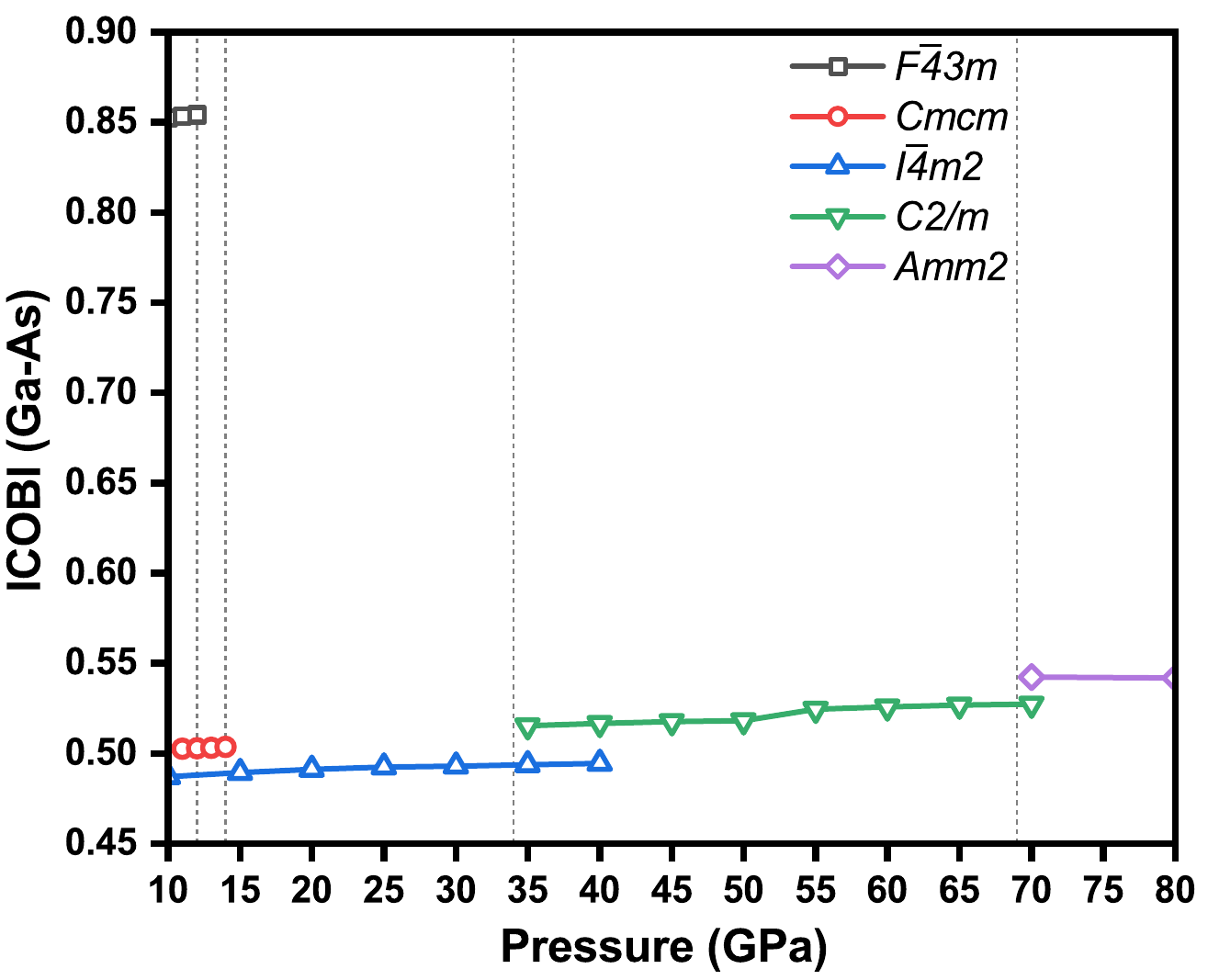}
	\end{minipage}
	\hfill
	\begin{minipage}{0.30\textwidth}
		\makebox[\linewidth][l]{\textbf{(b)}}\\[1mm]
		\centering
		\includegraphics[width=\linewidth]{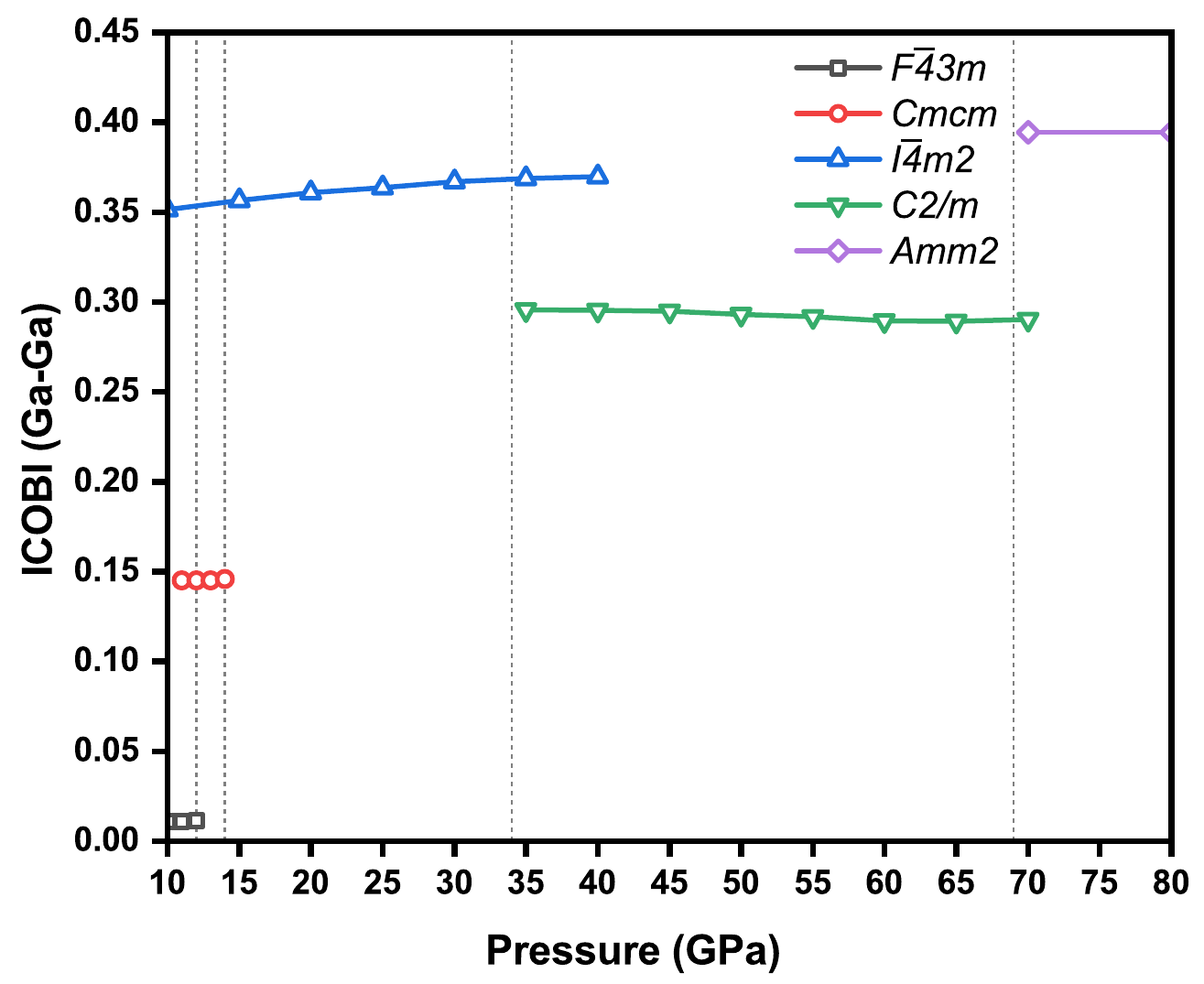}
	\end{minipage}
	\hfill
	\begin{minipage}{0.30\textwidth}
		\makebox[\linewidth][l]{\textbf{(c)}}\\[1mm]
		\centering
		\includegraphics[width=\linewidth]{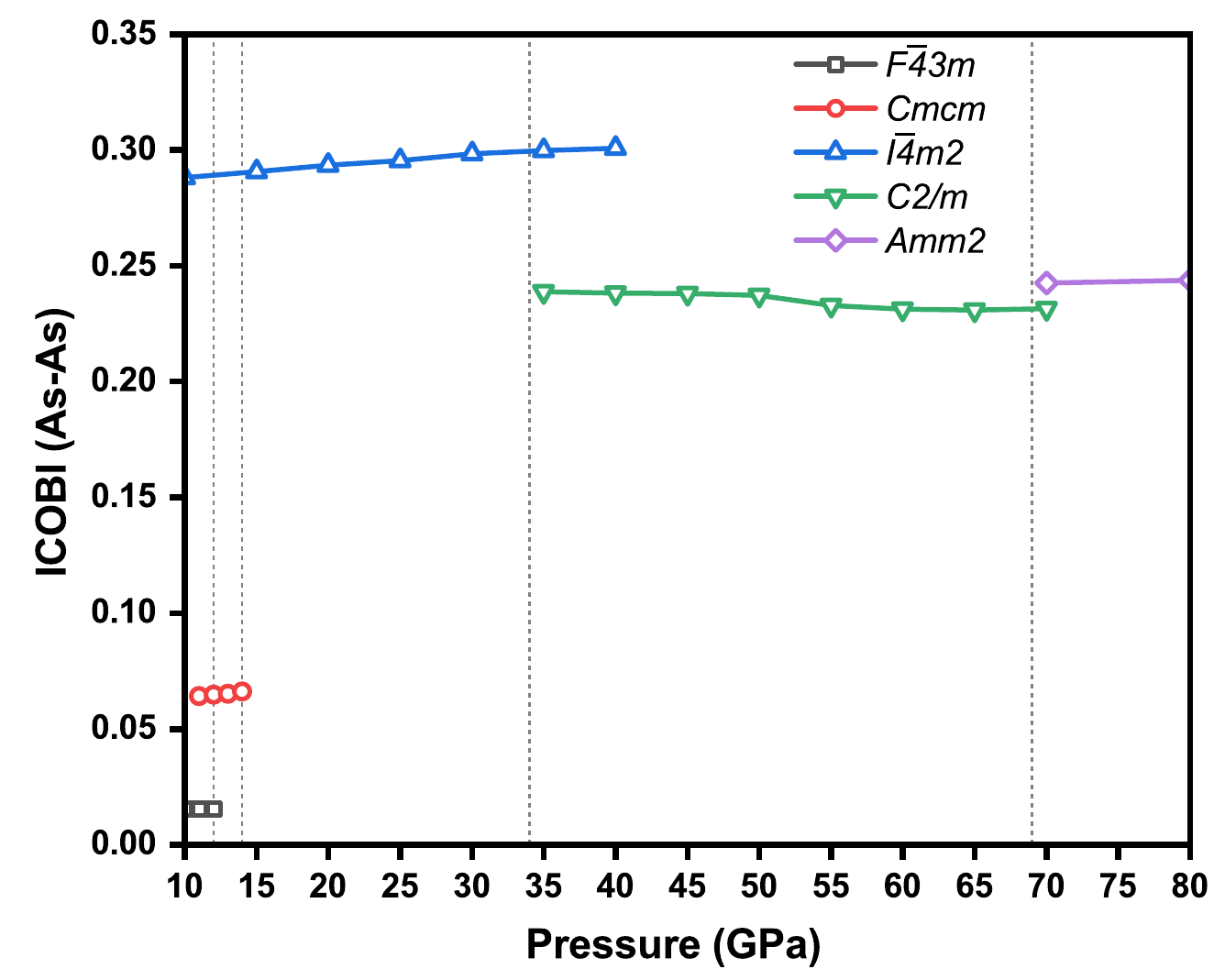}
	\end{minipage}
	
	\vspace{3mm}
	
	\begin{minipage}{0.30\textwidth}
		\makebox[\linewidth][l]{\textbf{(d)}}\\[1mm]
		\centering
		\includegraphics[width=\linewidth]{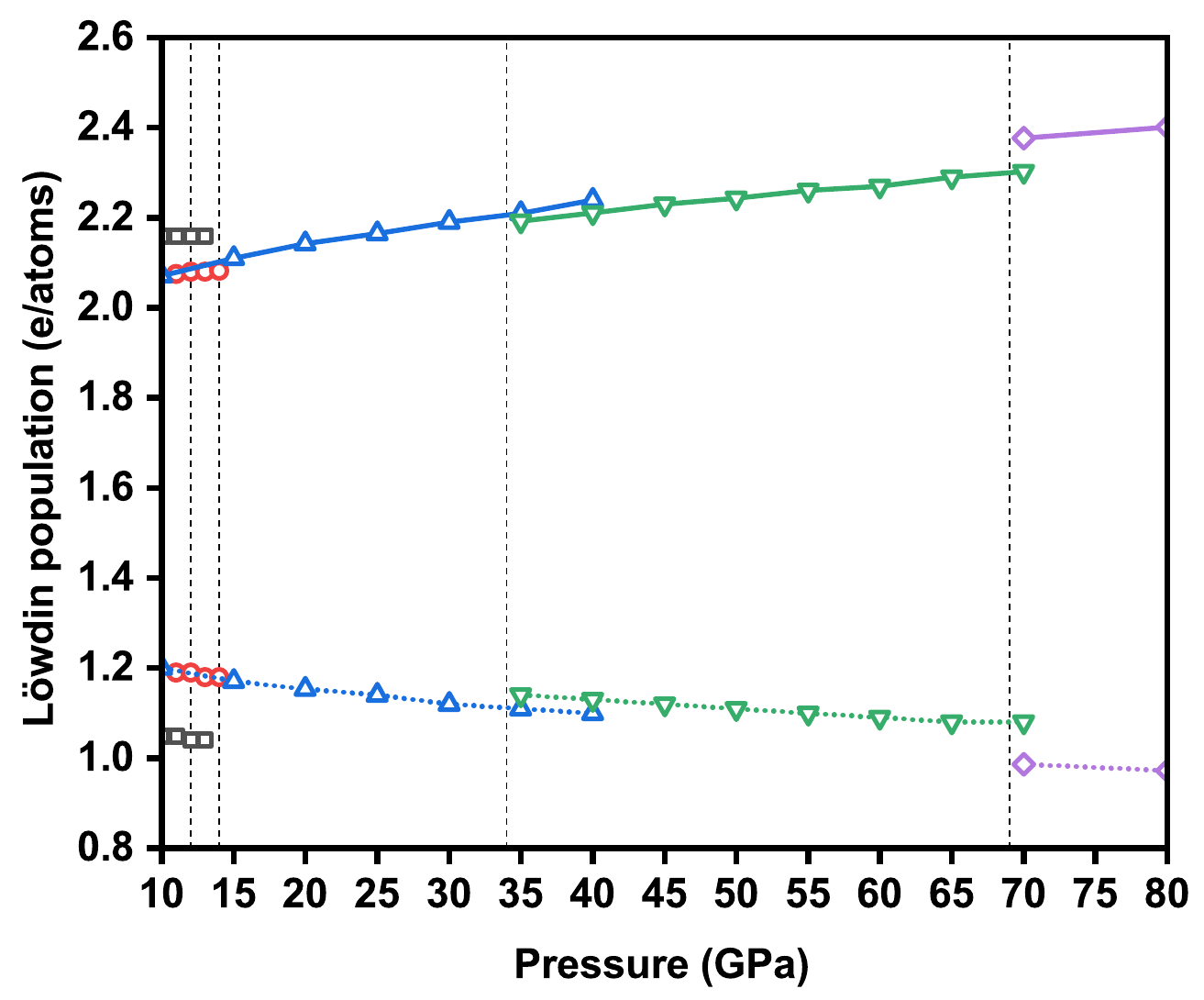}
	\end{minipage}
	\hspace{0.05\textwidth}
	\begin{minipage}{0.30\textwidth}
		\makebox[\linewidth][l]{\textbf{(e)}}\\[1mm]
		\centering
		\includegraphics[width=\linewidth]{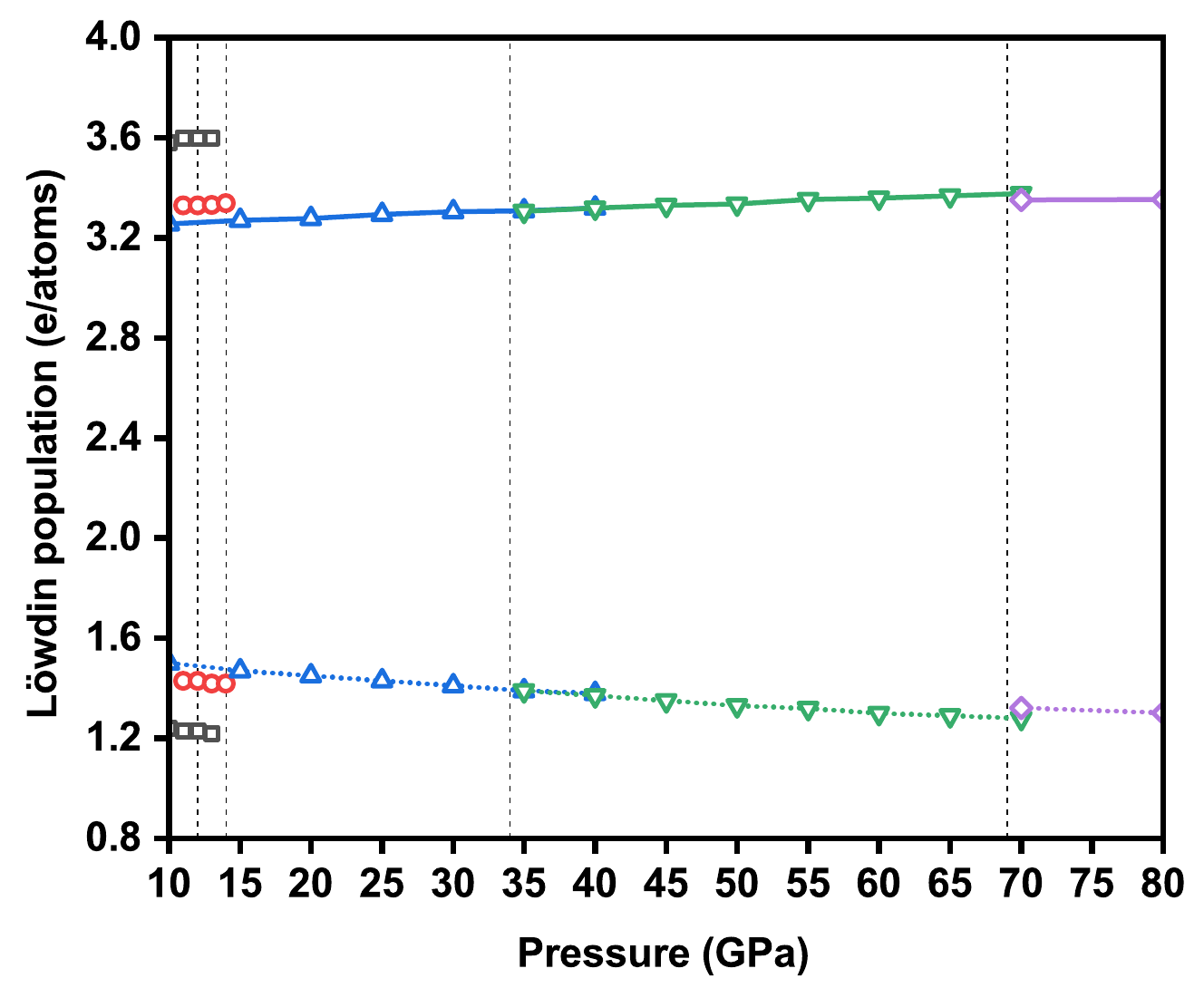}
	\end{minipage}

	\caption{Pressure dependent bonding characteristics and L\"owdin population in predicted GaAs phases. (a-c) shows the ICOBI values of (a) Ga-As (b) Ga-Ga (c) As-As bonding. (d-e) shows the L\"owdin population (d) for Ga atoms and (e) for As atoms for all GaAs phases for studied pressure range. In the plots vertical dashed line represent respective transition pressure. In figure (d and e), dot and solid connections within data points represents $4s$ and $4p$ orbitals, respectively.}
	\label{fig:lobster}
\end{figure*}

The nearest-contact Integrated Crystal Orbital Bond Index (ICOBI) values reveal a structure-dependent redistribution of electron sharing under compression. The ZB phase is characterized by a Ga--As ICOBI near 0.85 and very small homonuclear indices. The $Cmcm$ phase combines a lower nearest Ga--As index, close to 0.50, with appreciably larger Ga--Ga and As--As contributions. The $I\bar{4}m2$ phase combines a nearest Ga--As index around 0.49 with comparatively large Ga--Ga and As--As indices, all of which increase over its investigated pressure range. The $C2/m$ phase instead shows a nearest Ga--As index of approximately 0.52, which rises under compression, alongside weakly decreasing homonuclear indices overall (Fig. \ref{fig:lobster}(a--c)). These contrasting trends emphasize the dependence of bond-order character on both atomic arrangement and contact type. The electronic band gap can be correlated here, as the homonuclear electronic sharing increases system becomes metallic. The Löwdin populations likewise indicate changes in the distribution of occupied orbital character. Relative to $F\bar{4}3m$ at the same pressure of 12 GPa, $Cmcm$ assigns more electrons to the Ga and As 4s orbitals and fewer to their combined 4p orbitals. Compression within $I\bar{4}m2$ and $C2/m$ subsequently produces decreasing 4s populations together with increasing 4p populations for both elements. The small changes in the 3d populations suggest that most of the observed population redistribution involves the valence s and p orbitals (Fig. \ref{fig:lobster}(d and e)).

In conclusion, this study investigated the pressure-induced structural evolution of GaAs using first-principles DFT calculations together with a pressure--temperature phase diagram derived within the QHA. This approach extends the static enthalpy analysis by incorporating vibrational contributions to the Helmholtz free energy, enabling an assessment of phase stability at pressures up to 80 GPa and temperatures up to 1000 K. The enthalpy comparisons reveal a sequence of structural transformations from the ambient-pressure ZB  to $Cmcm$, $I\bar{4}m2$, $C2/m$, and finally $Amm2$, with transition pressures of approximately 12, 14, 34, and 69 GPa, respectively. The previously proposed $Imm2$ structure lies above the calculated minimum-enthalpy envelope. This sequence involves changes in crystal symmetry and atomic arrangement, reflecting the distinct structural responses of GaAs to compression. The absence of imaginary phonon frequencies in the calculated dispersions supports the dynamical stability of the investigated structures at the examined pressures. The calculated elastic constants satisfy the applied stability criteria, supporting mechanical stability under the corresponding conditions. The band structures indicate a change from the semiconducting behaviour of the ambient-pressure to metallic behaviour in all the predicted high-pressure phases. The resulting phase diagram demonstrates that temperature influences the relative stability of the predicted phases. This study provides a revised picture of high-pressure phase stability in GaAs by identifying competing phases and their electronic and vibrational properties.

\begin{acknowledgments}
	Acknowledgments -- 	The authors acknowledge the National Supercomputing Mission (NSM) for providing computational resources through PARAM Porul at the National Institute of Technology Tiruchirappalli (NIT Trichy), India. Some of the calculations were performed using the high-performance computing (HPC) facility established under the Anusandhan National Research Foundation (ANRF)–Partnerships for Accelerated Innovation and Research (PAIR) project (ANRF/PAIR/2025/000017/PAIR) at MSU Baroda, as well as at the Poznań Supercomputing and Networking Center (PSNC) under Grant No. pl0331-03. SP gratefully acknowledges Prof. Prafulla K. Jha for providing the opportunity to work on this problem during an internship.

\end{acknowledgments}

\bibliography{ref}
\end{document}


\title{\textbf{Supplemental Material\\ of \\ Mapping phase competition in GaAs under pressure and temperature} 
}
\author{Sakshi Patel}
\email{sakshipatelr23@gmail.com}  
\affiliation{Department of Physics, Faculty of Science, The Maharaja Sayajirao University of Baroda, Vadodara-390002, Gujarat, India.}

\author{Paras Patel}
\email{pparas727@gmail.com}  
\affiliation{Department of Physics, Faculty of Science, The Maharaja Sayajirao University of Baroda, Vadodara-390002, Gujarat, India.}

\author{Prafulla K. Jha}
\email{prafullaj@yahoo.com}  
\affiliation{Department of Physics, Faculty of Science, The Maharaja Sayajirao University of Baroda, Vadodara-390002, Gujarat, India.}  
\maketitle

\section{\label{I}Computational Details}

Candidate crystal structures of GaAs were explored at 0, 10, 20, 50, and 80 GPa using the particle-swarm optimization method implemented in CALYPSO \cite{PSO,CALYPSO}. Local structural relaxations and subsequent first-principles calculations were performed within density functional theory (DFT) using the Vienna \textit{Ab initio} Simulation Package (VASP, version 6.4.2) \cite{VASP}. Exchange and correlation were described using the PBEsol functional \cite{PBEsol}. The interactions between valence electrons and ionic cores were treated using the projector augmented-wave (PAW) method \cite{PAW}. The Ga $3d^{10}4s^{2}4p^{1}$ and As $3d^{10}4s^{2}4p^{3}$ electronic configurations were explicitly included in the valence treatment. The predicted structures and previously proposed high-pressure phases were relaxed over 0--80 GPa in increments of 5 GPa. At each pressure, the atomic coordinates, cell shape, and cell volume were optimized under the imposed hydrostatic pressure. A plane-wave kinetic energy cutoff of 700 eV was employed. Preliminary pressure dependent relaxations using conventional unit cells exhibited small numerical irregularities in the energy and enthalpy curves. These irregularities coincided with discrete changes in the automatically generated $k$-point meshes as the lattice contracted under compression. To reduce these sampling related effects, subsequent structural relaxations and property calculations were performed using supercell representations of the investigated phases. With the adopted supercells and reciprocal-space resolution of $0.03\times2\pi~\mathrm{\AA}^{-1}$, smoother pressure-dependent energy and enthalpy curves were obtained over 0--80 GPa. The electronic convergence threshold was $10^{-6}$ eV, and structural relaxations were continued until the residual atomic forces were below $0.01~\mathrm{eV\,\AA}^{-1}$. Gaussian smearing with a width of 0.2 eV was used. Non-spherical contributions to the gradient corrections within the PAW spheres were included explicitly. For the selected phases, lattice dynamics were investigated using the finite-displacement method implemented in Phonopy \cite{Phonopy}. Atomic forces calculated with VASP for displaced supercells were used to construct the harmonic force constants and obtain the phonon dispersion relations. Temperature-dependent free energies were additionally evaluated within the quasiharmonic approximation (QHA) using Phonopy. In this approach, harmonic phonon calculations at multiple volumes account for the volume dependence of the vibrational free energy, including zero-point and thermal contributions. Crystal structures were visualized using VESTA \cite{vesta}, and post-processing was performed using VASPKIT \cite{vaspkit} software. The symmetry related analysis is done using the FINDSYM \cite{findsym} utility.

\section{\label{II}Supporting Equations}

\subsection{Born Mechanical Stability Criteria}

For the cubic $F\bar{4}3m$ phase, the mechanical stability conditions are
\begin{equation}
C_{11}-C_{12}>0,\qquad
C_{11}+2C_{12}>0,\qquad
C_{44}>0.
\label{eq:cubic_stability}
\end{equation}

For the orthorhombic $Cmcm$ and $Amm2$ phases, the Born stability criteria are
\begin{equation}
\begin{aligned}
C_{11} &> 0, \qquad C_{11}C_{22} > C_{12}^{2},\\
C_{11}C_{22}C_{33} + 2C_{12}C_{13}C_{23}
&- C_{11}C_{23}^{2} - C_{22}C_{13}^{2} - C_{33}C_{12}^{2} > 0,\\
C_{44} &> 0, \qquad C_{55} > 0, \qquad C_{66} > 0.
\end{aligned}
\label{eq:orthorhombic_stability}
\end{equation}

For the tetragonal $I\bar{4}m2$ phase, the stability conditions are
\begin{equation}
\begin{aligned}
C_{11} &> |C_{12}|; \qquad
2C_{13}^{2} < C_{33}(C_{11}+C_{12})\\
C_{44} &> 0; \qquad C_{66} > 0
\end{aligned}
\label{eq:stability}
\end{equation}

For the monoclinic $C2/m$ phase, the Born stability criteria are expressed as
\begin{equation}
\begin{gathered}
C_{ii}>0,\quad \forall\,i\in\{1,\ldots,6\},\\
C_{11}+C_{22}+C_{33}+2(C_{12}+C_{13}+C_{23})>0,\\
C_{33}C_{55}-C_{35}^{2}>0,\\
C_{44}C_{66}-C_{46}^{2}>0,\\
C_{22}+C_{33}-2C_{23}>0,\\
C_{22}(C_{33}C_{55}-C_{35}^{2})+2C_{23}C_{25}C_{35}-C_{23}^{2}C_{55}-C_{25}^{2}C_{33}>0,\\
2[C_{15}C_{25}(C_{33}C_{12}-C_{13}C_{23})+C_{15}C_{35}(C_{22}C_{13}-C_{12}C_{23})+C_{25}C_{35}(C_{11}C_{23}-C_{12}C_{13})]\\
-[C_{15}^{2}(C_{22}C_{33}-C_{23}^{2})+C_{25}^{2}(C_{11}C_{33}-C_{13}^{2})+C_{35}^{2}(C_{11}C_{22}-C_{12}^{2})]
+C_{55}g>0.
\end{gathered}
\label{eq:monoclinic_stability}
\end{equation}

where
\begin{equation}
\begin{gathered}
g=C_{11}C_{22}C_{33}-C_{11}C_{23}^{2}
-C_{22}C_{13}^{2}-C_{33}C_{12}^{2}+2C_{12}C_{13}C_{23}.
\end{gathered}
\label{eq:g_definition}
\end{equation}

\clearpage
\section{\label{III}Supporting Table}
\subsection{Structural parameters of considered GaAs phases}
\begin{table*}[h]
\caption{Optimized lattice parameters and atomic positions of GaAs
phases at representative pressures.}
\label{tab:structural_parameters}
\centering
\scriptsize
\setlength{\tabcolsep}{4pt}
\renewcommand{\arraystretch}{1.4}

\begin{tabular}{c c c c}
\hline\hline
Phase (space group) &
Pressure (GPa) &
Lattice parameters (unit cell) &
Atom positions (Wyckoff) \\
\hline
\noalign{\vskip 3pt}

\makecell[c]{$F\bar{4}3m$\\(216)\\PBEsol} &
0 &
\makecell[c]{$a=b=c=5.659$ \AA{}\\
$\alpha=\beta=\gamma=90^\circ$} &
\makecell[c]{Ga ($4a$) 0.00000 0.00000 0.00000\\
As ($4c$) 0.25000 0.25000 0.25000} \\
\noalign{\vskip 3pt}
\hline
\noalign{\vskip 3pt}

\makecell[c]{$F\bar{4}3m$\\(216)\\Experiment \cite{Vurgaftman2001}} &
0 &
\makecell[c]{$a=b=c=5.653$ \AA{}\\
$\alpha=\beta=\gamma=90^\circ$} &
\makecell[c]{Ga ($4a$) 0.00000 0.00000 0.00000\\
As ($4c$) 0.25000 0.25000 0.25000} \\
\noalign{\vskip 3pt}
\hline
\noalign{\vskip 3pt}

\makecell[c]{$F\bar{4}3m$\\(216)\\GGA-PBE \cite{lattice_ZB_Theory}} &
0 &
\makecell[c]{$a=b=c=5.648$ \AA{}\\
$\alpha=\beta=\gamma=90^\circ$} &
\makecell[c]{Ga ($4a$) 0.00000 0.00000 0.00000\\
As ($4c$) 0.25000 0.25000 0.25000} \\
\noalign{\vskip 3pt}
\hline
\noalign{\vskip 3pt}

\makecell[c]{$Cmcm$\\(63)\\PBEsol} &
12 &
\makecell[c]{$a=5.066$ \AA{}, $b=5.326$ \AA{}\\
$c=4.876$ \AA{}\\
$\alpha=\beta=\gamma=90^\circ$} &
\makecell[c]{Ga ($4c$) 0.00000 0.33649 0.75000\\
As ($4c$) 0.00000 0.81902 0.75000} \\
\noalign{\vskip 3pt}
\hline
\noalign{\vskip 3pt}

\makecell[c]{$Cmcm$\\(63)\\PBEsol} &
20 &
\makecell[c]{$a=4.976$ \AA{}, $b=5.275$ \AA{}\\
$c=4.727$ \AA{}\\
$\alpha=\beta=\gamma=90^\circ$} &
\makecell[c]{Ga ($4c$) 0.00000 0.349902 0.75000\\
As ($4c$) 0.00000 0.831427 0.75000} \\
\noalign{\vskip 3pt}
\hline
\noalign{\vskip 3pt}

\makecell[c]{$Cmcm$\\(63)\\Experiment \cite{McMahon_Cmcm}} &
22.7 &
\makecell[c]{$a=4.960$ \AA{}, $b=5.290$ \AA{}\\
$c=4.819$ \AA{}\\
$\alpha=\beta=\gamma=90^\circ$} &
\makecell[c]{Ga ($4c$) 0.00000 0.64500 0.75000\\
As ($4c$) 0.00000 0.16400 0.75000} \\
\noalign{\vskip 3pt}
\hline
\noalign{\vskip 3pt}

\makecell[c]{$I\bar{4}m2$\\(119)} &
15 &
\makecell[c]{$a=b=4.885$ \AA{}\\
$c=2.675$ \AA{}\\
$\alpha=\beta=\gamma=90^\circ$} &
\makecell[c]{Ga ($2c$) 0.00000 0.50000 0.25000\\
As ($2b$) 0.00000 0.00000 0.50000} \\
\noalign{\vskip 3pt}
\hline
\noalign{\vskip 3pt}

\makecell[c]{$C2/m$\\(12)} &
35 &
\makecell[c]{$a=5.144$ \AA{}, $b=4.870$ \AA{}\\
$c=2.650$ \AA{}\\
$\alpha=\gamma=90^\circ$, $\beta=121.003^\circ$} &
\makecell[c]{Ga ($2b$) 0.00000 0.50000 0.00000\\
As ($2a$) 0.00000 0.00000 0.00000} \\
\noalign{\vskip 3pt}
\hline
\noalign{\vskip 3pt}

\makecell[c]{$Amm2$\\(38)} &
70 &
\makecell[c]{$a=4.646$ \AA{}, $b=12.599$ \AA{}\\
$c=4.274$ \AA{}\\
$\alpha=\beta=\gamma=90^\circ$} &
\makecell[c]{Ga ($2a$) 0.00000 0.00000 0.49907\\
Ga ($4d$) 0.00000 0.09578 0.99918\\
Ga ($4e$) 0.50000 0.20133 0.50085\\
As ($2b$) 0.50000 0.00000 0.50079\\
As ($4d$) 0.00000 0.19377 0.49932\\
As ($4e$) 0.50000 0.39248 0.50072} \\
\noalign{\vskip 3pt}
\hline\hline

\end{tabular}
\end{table*}

\clearpage

\subsection{Raman and IR active mode of $I\bar{4}m2$, $C2/m$ and $Amm2$}
\begin{table*}[h]
	\caption{
		Zone-center optical phonon frequencies and their Raman and infrared (IR)
		activities for the $I\bar{4}m2$, $C2/m$ and  $Amm2$ phases.
		Only optical modes are listed; the three acoustic modes are excluded.
	}
	\label{tab:raman_ir_modes}
	\centering
	\small
	\setlength{\tabcolsep}{5pt}
	\renewcommand{\arraystretch}{1.25}
	
	\begin{tabular}{c c l c c}
		\hline\hline
		Phase &
		Irrep &
		\multicolumn{1}{c}{Frequency (cm$^{-1}$)} &
		Raman &
		IR \\
		\hline
		
		$I\bar{4}m2$ &
		$B_{2}$ &
		69.11 &
		Yes &
		Yes \\
		
		(15 GPa)&
		$E$ &
		223.45 &
		Yes &
		Yes \\
		
		\hline
		
		$C2/m$ &
		$A_{u}$ &
		277.36 &
		No &
		Yes \\
		
		(35 GPa)&
		$B_{u}$ &
		54.84, 238.06 &
		No &
		Yes \\
				\hline
		
		$Amm2$ &
		$A_{1}$ &
		57.77, 163.38, 185.16, 188.10, 211.85, 257.11, 283.06, 293.47, 328.83 &
		Yes &
		Yes \\
		
		(70 GPa)&
		$A_{2}$ &
		71.55, 78.02, 354.14, 368.66 &
		Yes &
		No \\
		
		&
		$B_{1}$ &
		80.86, 83.32, 353.41, 364.05, 399.24 &
		Yes &
		Yes \\
		
		&
		$B_{2}$ &
		57.21, 159.48, 172.69, 188.36, 207.18, 262.45, 282.33, 295.24, 334.70 &
		Yes &
		Yes \\
		
		\hline\hline
	\end{tabular}
\end{table*}

\clearpage

\subsection{Elastic constants and mechanical stability of GaAs phases}
\begin{table*}[h]
\caption{Lastic constants and polycrystalline elastic moduli of the high-pressure phases of GaAs at representative pressures.}
\label{tab:structural_parameters}
\centering
\scriptsize
\setlength{\tabcolsep}{4pt}
\renewcommand{\arraystretch}{1.4}

\begin{ruledtabular}
		\begin{tabular}{c c c c c c c}
			Phase (space group) &
			Pressure (GPa) &
			$C_{ij}$ (GPa) &
			$B$ (GPa) &
			$E$ (GPa) &
			$G$ (GPa) &
			$\nu$ \\
			\hline
			
			\makecell[c]{$F\bar{4}3m$\\(216)\\PBEsol} &
			0 &
			\makecell[c]{
				$C_{11}=109.61$\\
				$C_{12}=49.63$\\
				$C_{44}=54.11$
			} &
			69.62 & 106.36 & 42.70 & 0.25 \\
			
			\hline
			
			\makecell[c]{$F\bar{4}3m$\\(216)\\Experiment \cite{elastic_ZB_Exp}} &
			0 &
			\makecell[c]{
				$C_{11}=118.77$\\
				$C_{12}=53.72$\\
				$C_{44}=59.44$
			} &
			75.40 & -- & -- & -- \\
			
			\hline
			
			\makecell[c]{$F\bar{4}3m$\\(216)\\GGA (PW-91) \cite{elastic_ZB_Theory}} &
			0 &
			\makecell[c]{
				$C_{11}=108.3$\\
				$C_{12}=49.5$\\
				$C_{44}=57.6$
			} &
			69.1 & 108.8 & 44 & 0.24 \\
			
			\hline
			
			\makecell[c]{$Cmcm$\\(63)} &
			12 &
			\makecell[c]{
				$C_{11}=272.30,\quad C_{22}=170.61$\\
				$C_{33}=157.71$\\
				$C_{12}=58.97,\quad C_{13}=32.76$\\
				$C_{23}=125.57$\\
				$C_{44}=13.03,\quad C_{55}=42.85$\\
				$C_{66}=56.46$
			} &
			113.87 & 102.19 & 37.84 & 0.35 \\
			
			\hline
			
			\makecell[c]{$I\bar{4}m2$\\(119)} &
			15 &
			\makecell[c]{
				$C_{11}=178.28,\quad C_{12}=148.91$\\
				$C_{13}=68.79,\quad C_{33}=236.14$\\
				$C_{44}=44.92,\quad C_{66}=77.24$
			} &
			129.42 & 121.36 & 45.16 & 0.34 \\
			
			\hline
			
			\makecell[c]{$C2/m$\\(12)} &
			35 &
			\makecell[c]{
				$C_{11}=332.33,\quad C_{22}=436.00$\\
				$C_{33}=322.55$\\
				$C_{12}=56.90,\quad C_{13}=168.50$\\
				$C_{23}=69.71$\\
				$C_{15}=-16.28,\quad C_{25}=-13.86$\\
				$C_{35}=26.36$\\
				$C_{44}=72.11,\quad C_{46}=14.32$\\
				$C_{55}=44.00,\quad C_{66}=84.64$
			} &
			186.76 & 211.56 & 80.67 & 0.31 \\
			
			\hline
			
			\makecell[c]{$Amm2$\\(38)} &
			70 &
			\makecell[c]{
				$C_{11}=661.18,\quad C_{22}=460.33$\\
				$C_{33}=521.37$\\
				$C_{12}=123.71,\quad C_{13}=85.31$\\
				$C_{23}=238.14$\\
				$C_{44}=127.73,\quad C_{55}=95.97$\\
				$C_{66}=94.49$
			} &
			281.75 & 345.45 & 133.315 & 0.29 \\
			
		\end{tabular}
	\end{ruledtabular}
\end{table*}

\clearpage
\section{\label{IV}Supporting Figures}
\subsection{Enthalpy difference between $Imm2$ and $C2/m$}

\begin{figure*}[h]
	\centering
	
	\begin{minipage}{0.42\textwidth}
		\makebox[\linewidth][l]{\textbf{(a)}}\\[2mm]
		\includegraphics[width=\linewidth]{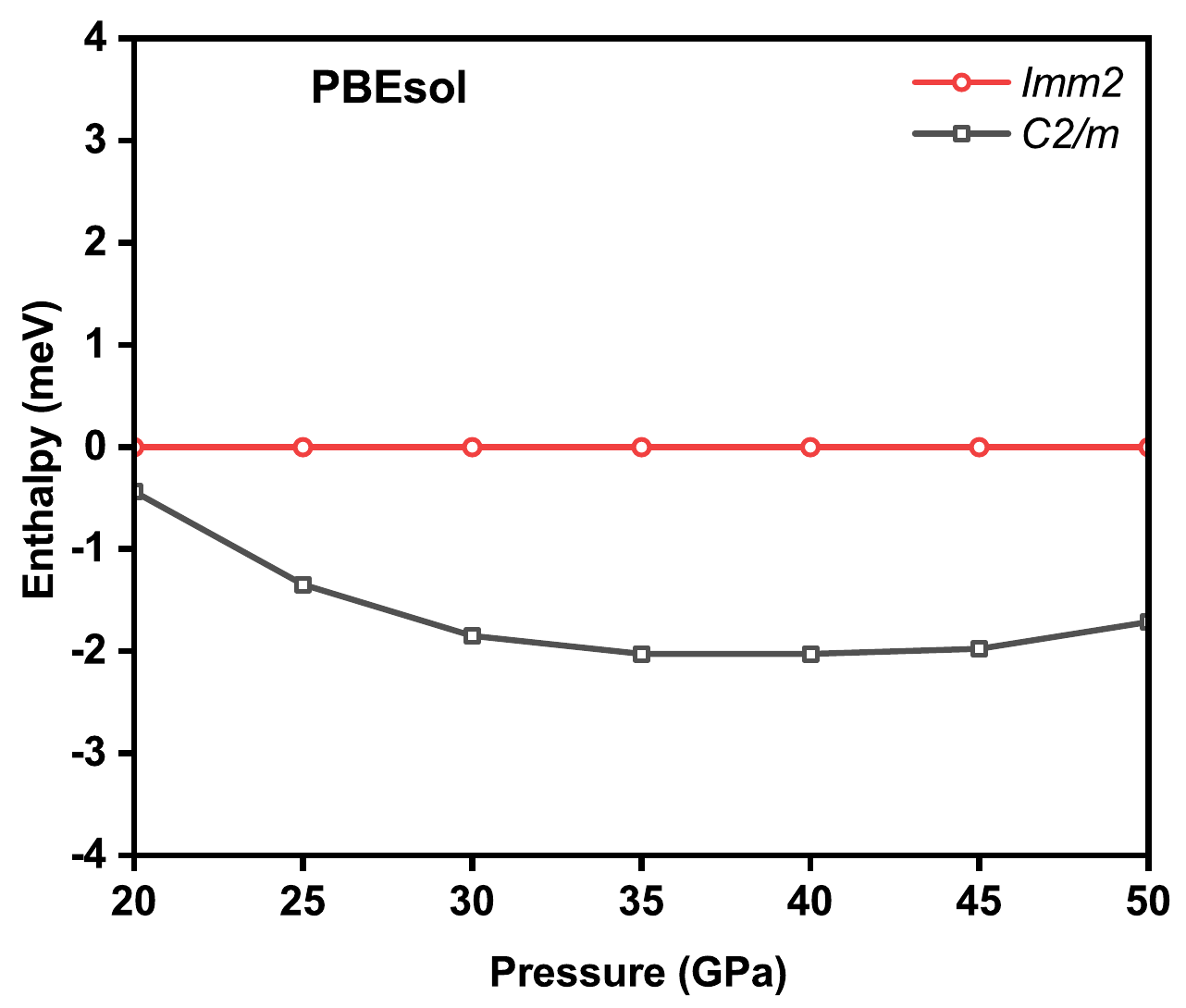}
	\end{minipage}
	\hfill
	\begin{minipage}{0.42\textwidth}
		\makebox[\linewidth][l]{\textbf{(}\textbf{b}\textbf{)}}\\[2mm]
		\includegraphics[width=\linewidth]{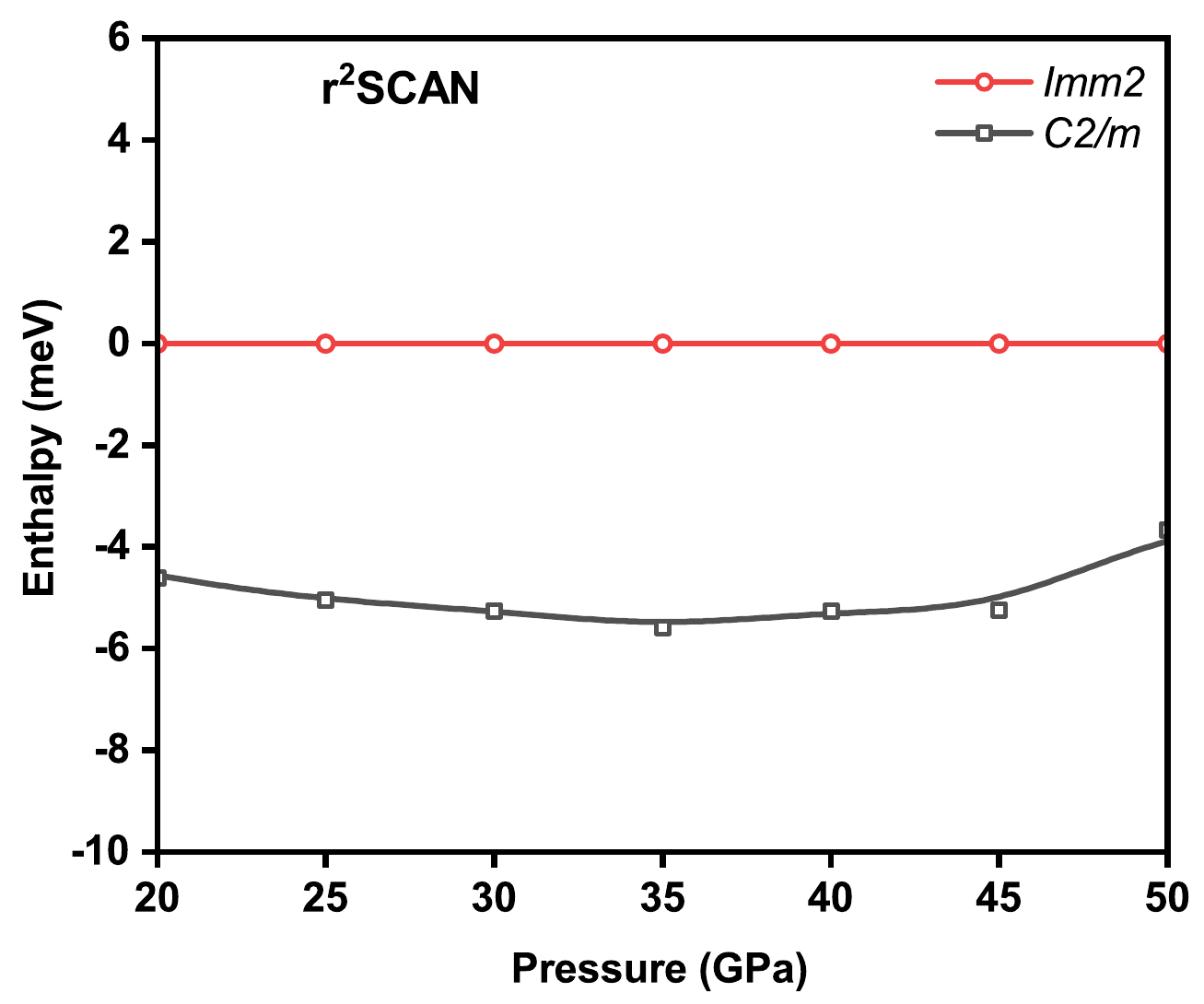}
	\end{minipage}
	\caption{Enthalpy curve for the two competing phases, $C2/m$ and $Imm2$ using (a) $PBEsol$ and (b) $r^2SCAN$ functional. Both calculations shows $C2/m$ as energetically favoured phase.}
	\label{fig:enthalpy}
\end{figure*}

\clearpage
\subsection{Phonon dispersion of $Imm2$ phase}

\begin{figure}[h]
	\centering
	
	\begin{minipage}{0.65\columnwidth} 
		\raggedright 
		{\fontsize{12}{14}\selectfont} \ 
		\includegraphics[width=\linewidth]{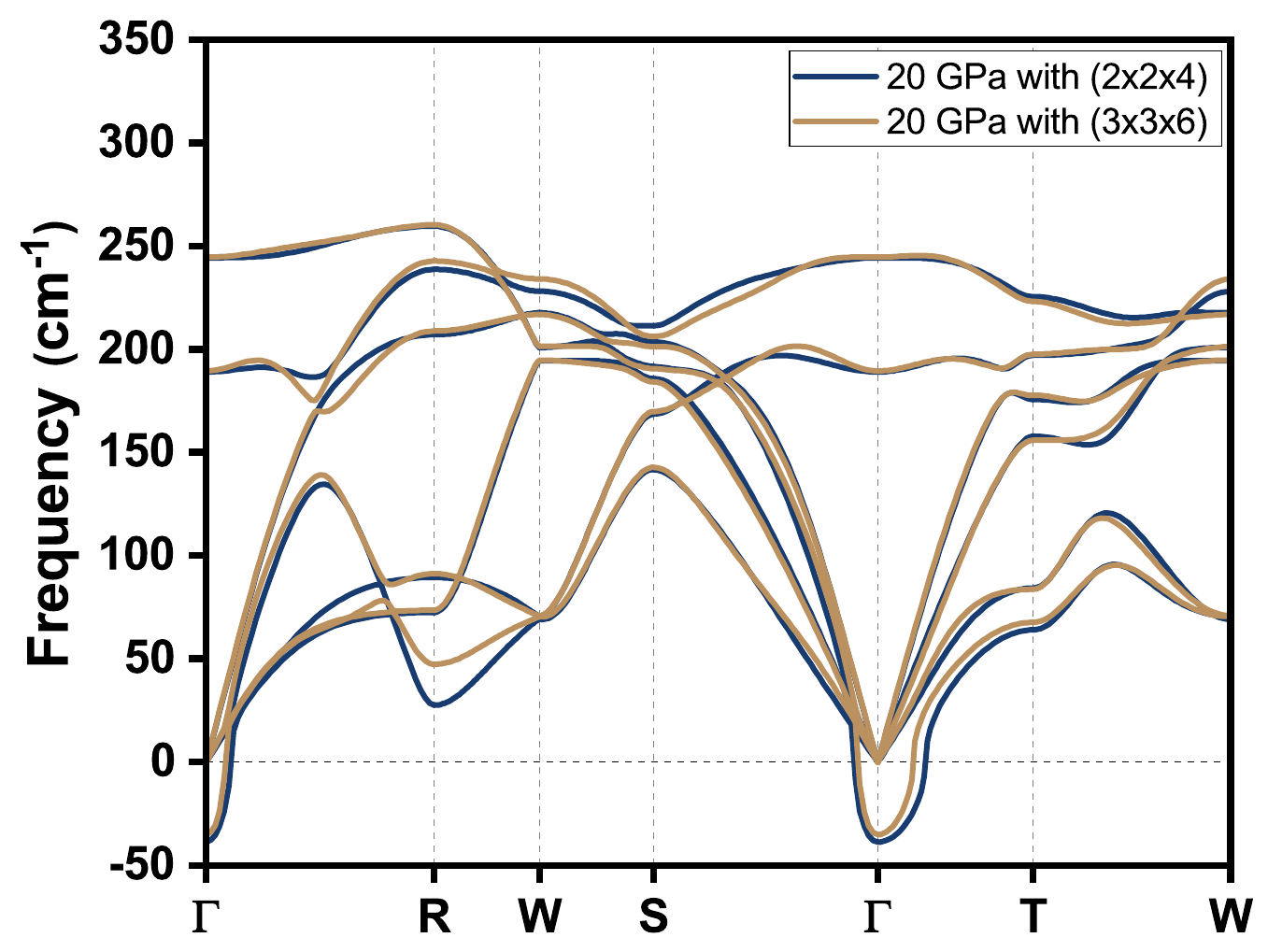} 
	\end{minipage}
	\caption{Phonon dispersion curve of $Imm2$ phase of GaAs at 20 GPa using two different supercell configuration $(2\times2\times4)$ (blue colour) and $(3\times3\times6)$ (brown colour). Both supercell calculation exhibits imaginary phonon frequencies. The persistence of this imaginary modes upon increasing the supercell size indicates that the observed dynamical instability is real, not a numerical artifact.}
	
	\label{fig:PhononImm2}
\end{figure}

\clearpage

\subsection{Phonon dispersion curves for all considered GaAs phases}

\begin{figure*}[h]
	\centering
	
	\begin{minipage}{0.32\textwidth}
		\makebox[\linewidth][l]{\textbf{(a)}}\\[2mm]
		\includegraphics[width=\linewidth]{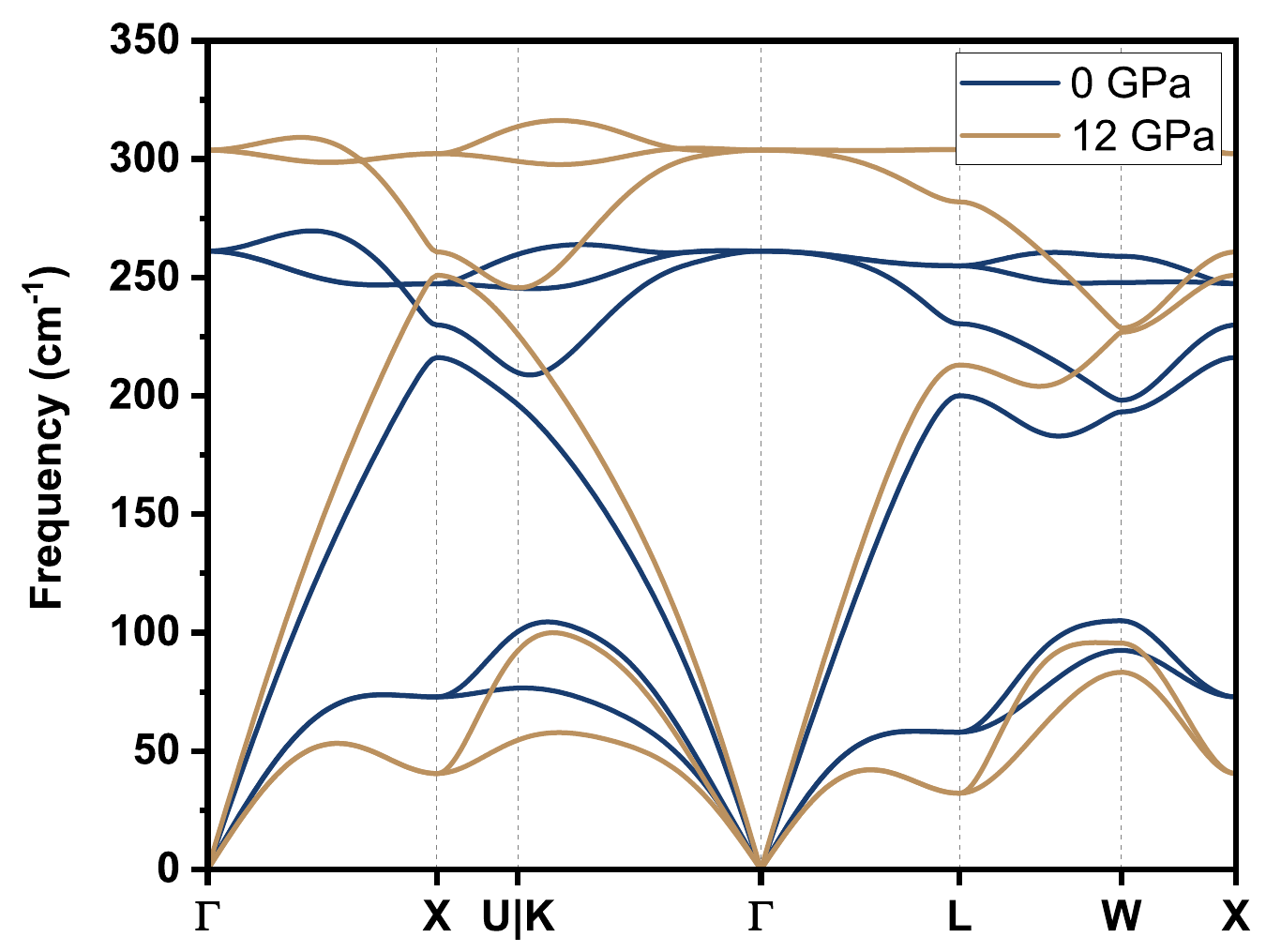}
	\end{minipage}
	\hfill
	\begin{minipage}{0.32\textwidth}
		\makebox[\linewidth][l]{\textbf{(}\textbf{b}\textbf{)}}\\[2mm]
		\includegraphics[width=\linewidth]{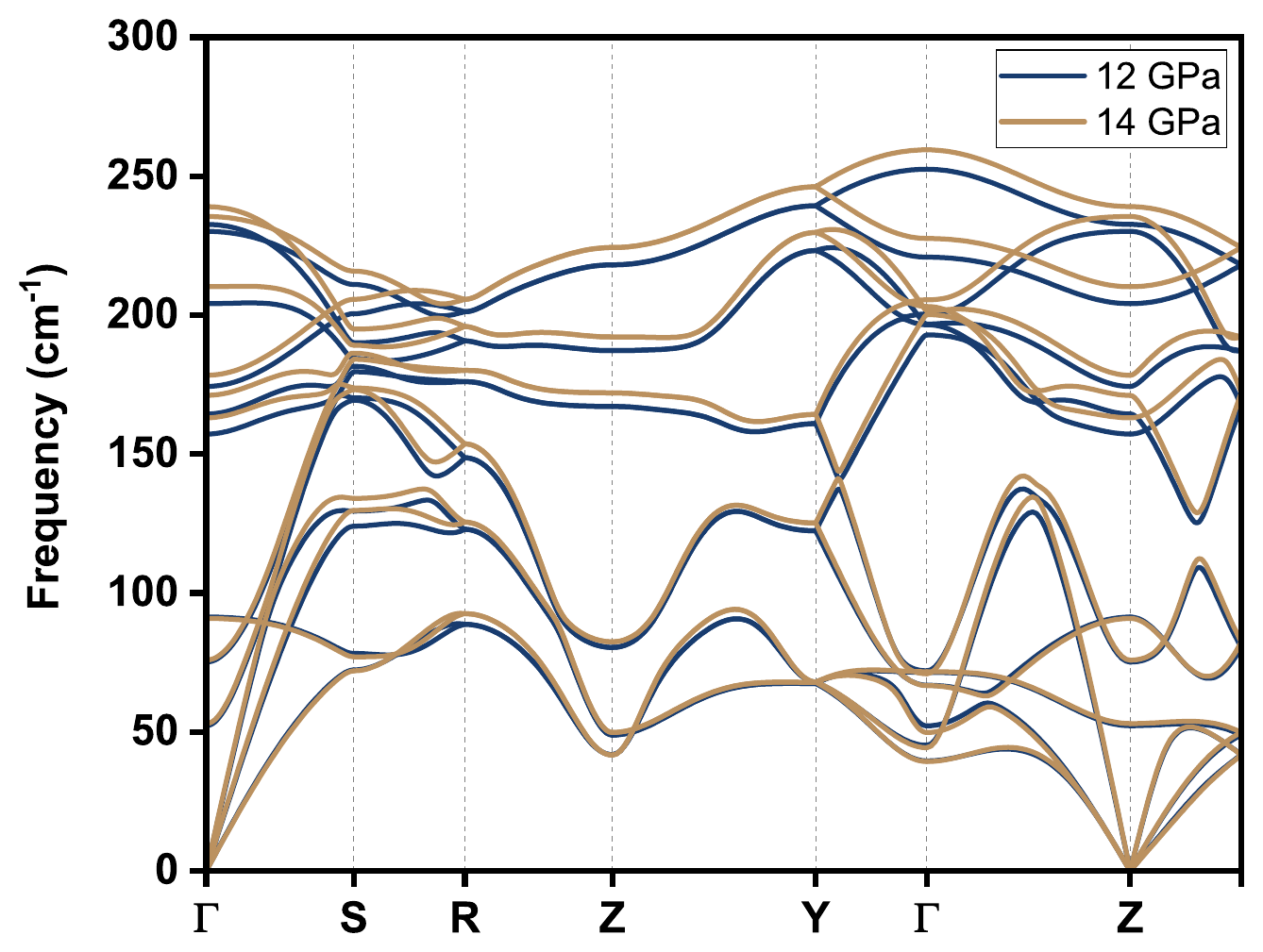}
	\end{minipage}
	\hfill
	\begin{minipage}{0.32\textwidth}
		\makebox[\linewidth][l]{\textbf{(}\textbf{c}\textbf{)}}\\[2mm]
		\includegraphics[width=\linewidth]{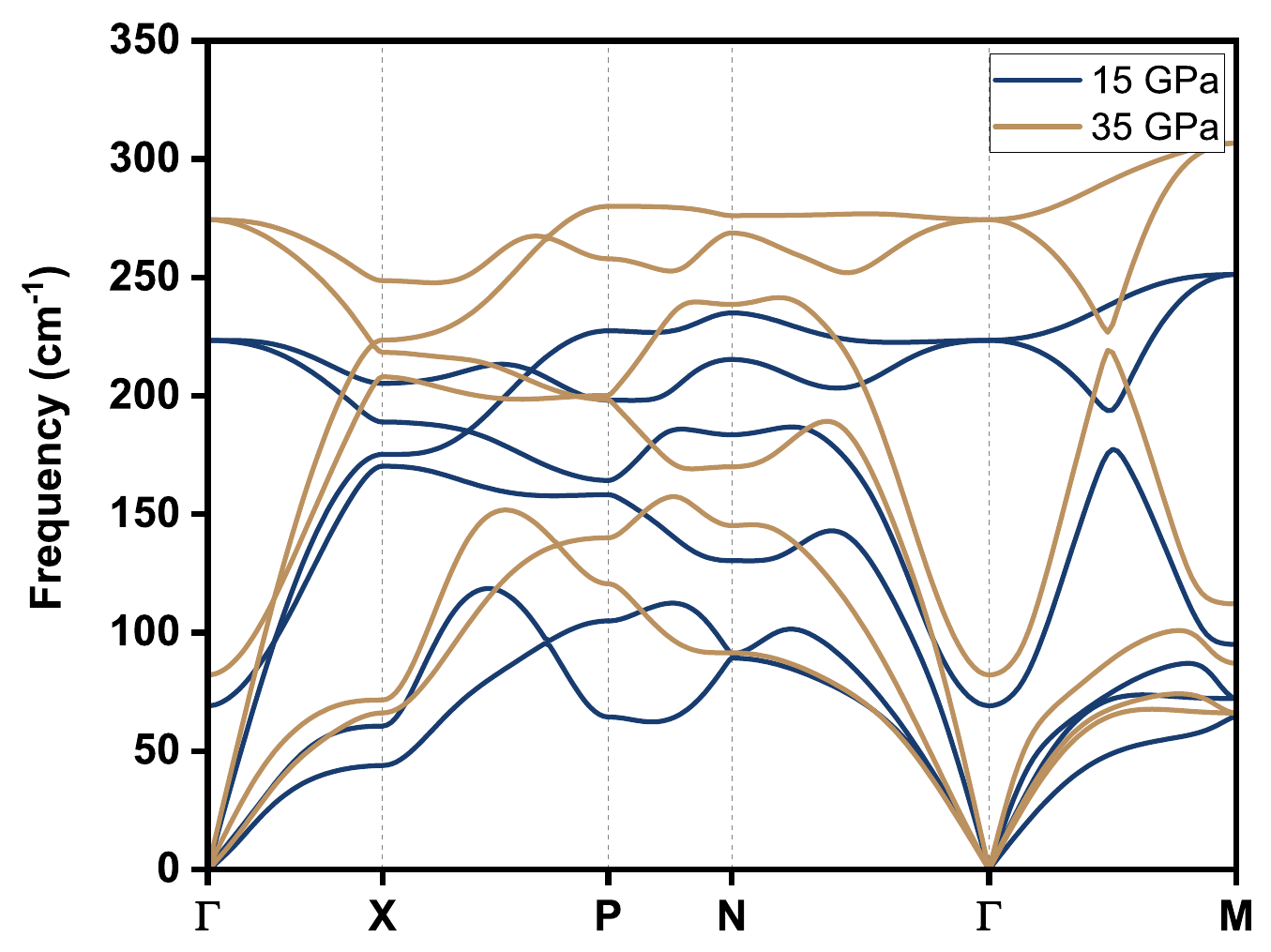}
	\end{minipage}
	\vfill
	\makebox[\textwidth][c]{%
		\begin{minipage}{0.32\textwidth}
			\makebox[\linewidth][l]{\textbf{(d)}}\\[2mm]
			\includegraphics[width=\linewidth]{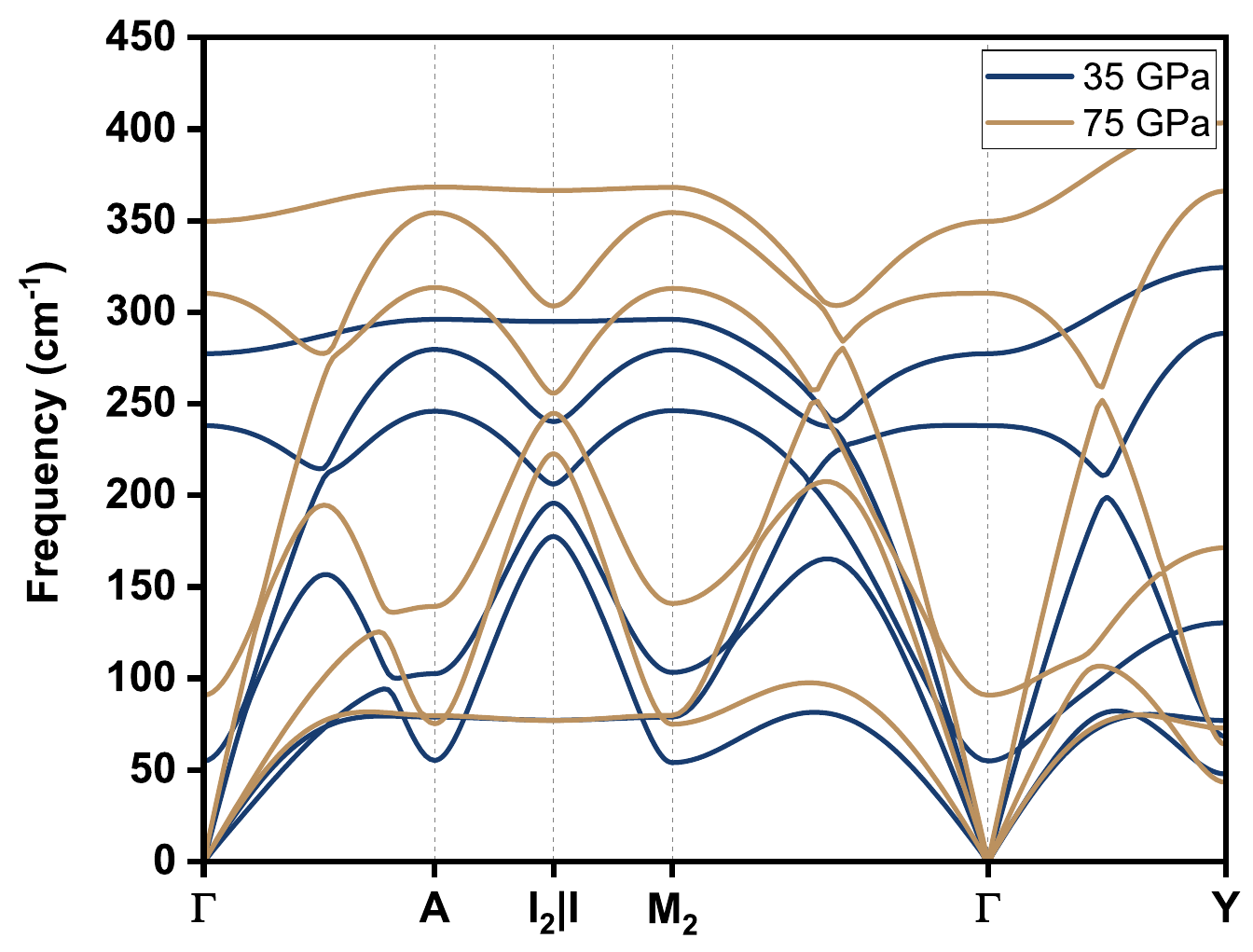}
		\end{minipage}
		\hspace{0.02\textwidth}
		\begin{minipage}{0.32\textwidth}
			\makebox[\linewidth][l]{\textbf{(e)}}\\[2mm]
			\includegraphics[width=\linewidth]{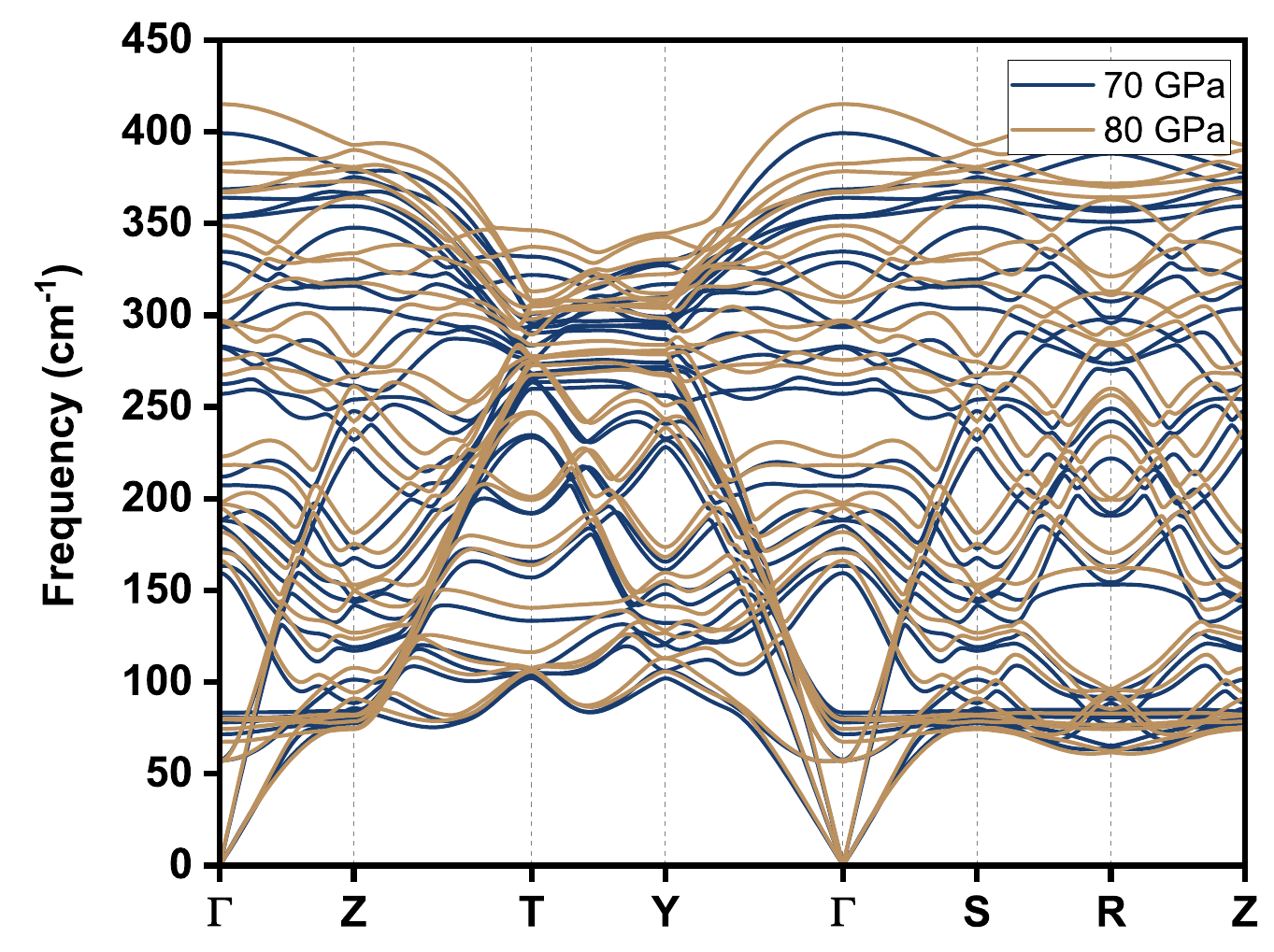}
		\end{minipage}%
	}
	
	\caption{Phonon dispersion curves of GaAs phases at energetically stable pressure boundaries, calculated using the finite-displacement method. The calculations employed $2\times2\times2$ supercells for the ZB, $Cmcm$, and $C2/m$ phases, a $3\times3\times5$ supercell for the $I\bar{4}m2$ phase, and a $3\times1\times3$ supercell for the $Amm2$ phase, ensuring supercell lattice dimensions greater than $10~\text{\AA}$. (a) $F\bar{4}3m$ at 0 and 12 GPa, (b) $Cmcm$ at 12 and 14 GPa, (c) $I\bar{4}m2$ at 15 and 35 GPa, (d) $C2/m$ at 35 and 75 GPa, and (e) $Amm2$ at 70 and 80 GPa.}
	\label{fig:phonon}
\end{figure*}
\clearpage

\subsection{Electronic band-structures}

\begin{figure*}[h]
	\centering
	
	\begin{minipage}{0.40\textwidth}
		\makebox[\linewidth][l]{\textbf{(a)}}\\[2mm]
		\includegraphics[width=\linewidth]{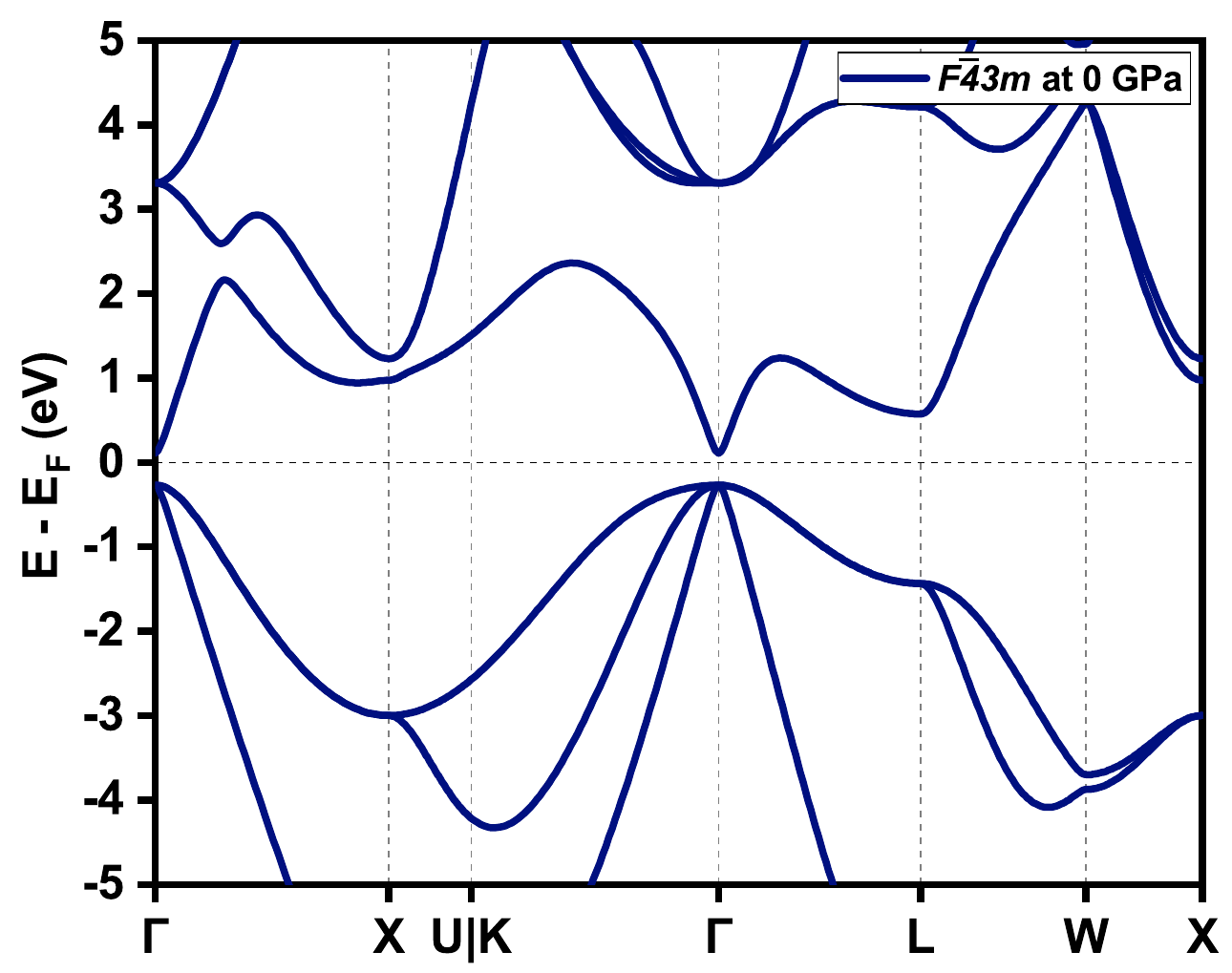}
	\end{minipage}
	\hfill
	\begin{minipage}{0.40\textwidth}
		\makebox[\linewidth][l]{\textbf{(}\textbf{b}\textbf{)}}\\[2mm]
		\includegraphics[width=\linewidth]{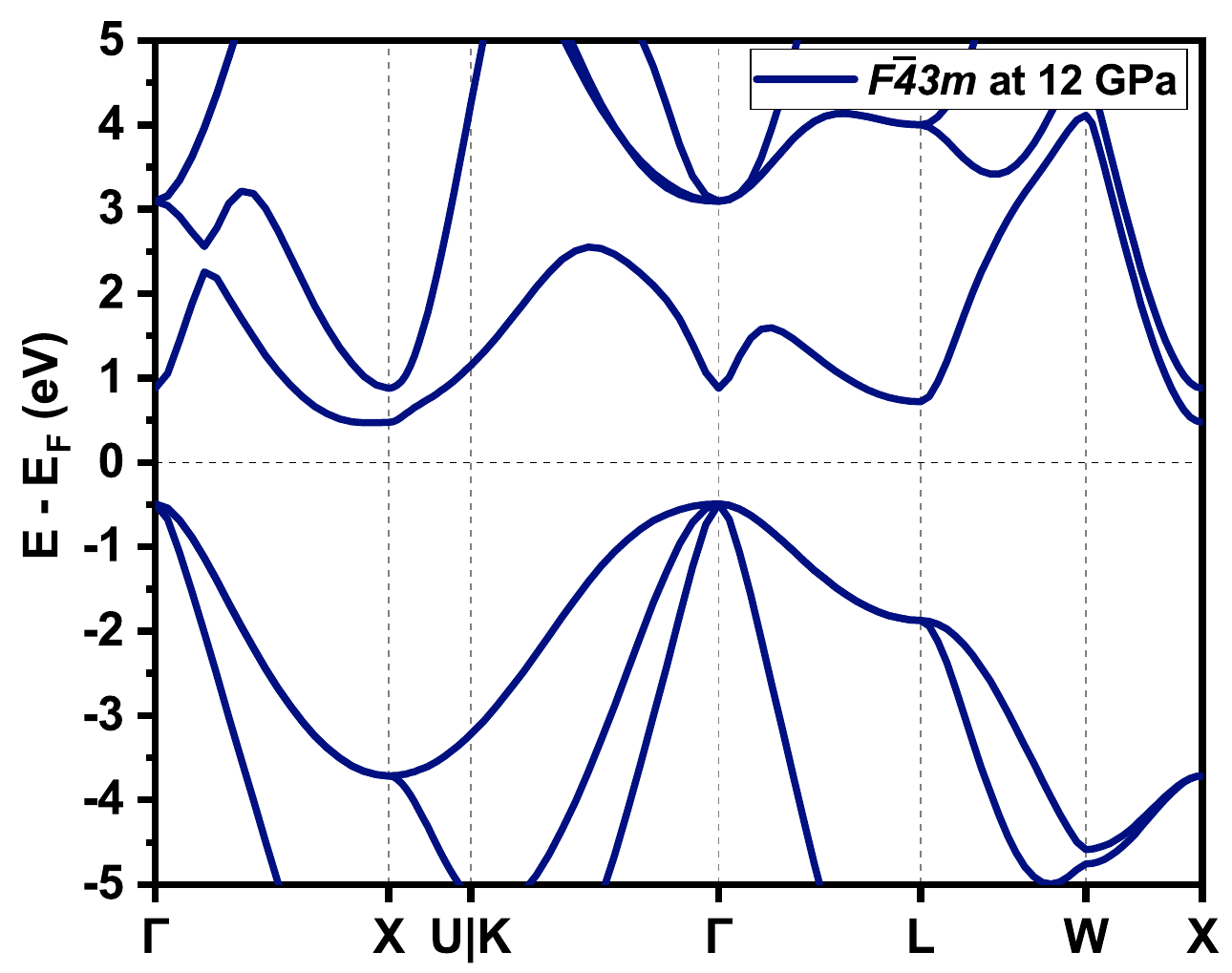}
	\end{minipage}
	
	\vfill
	\begin{minipage}{0.40\textwidth}
	\makebox[\linewidth][l]{\textbf{(c)}}\\[2mm]
	\includegraphics[width=\linewidth]{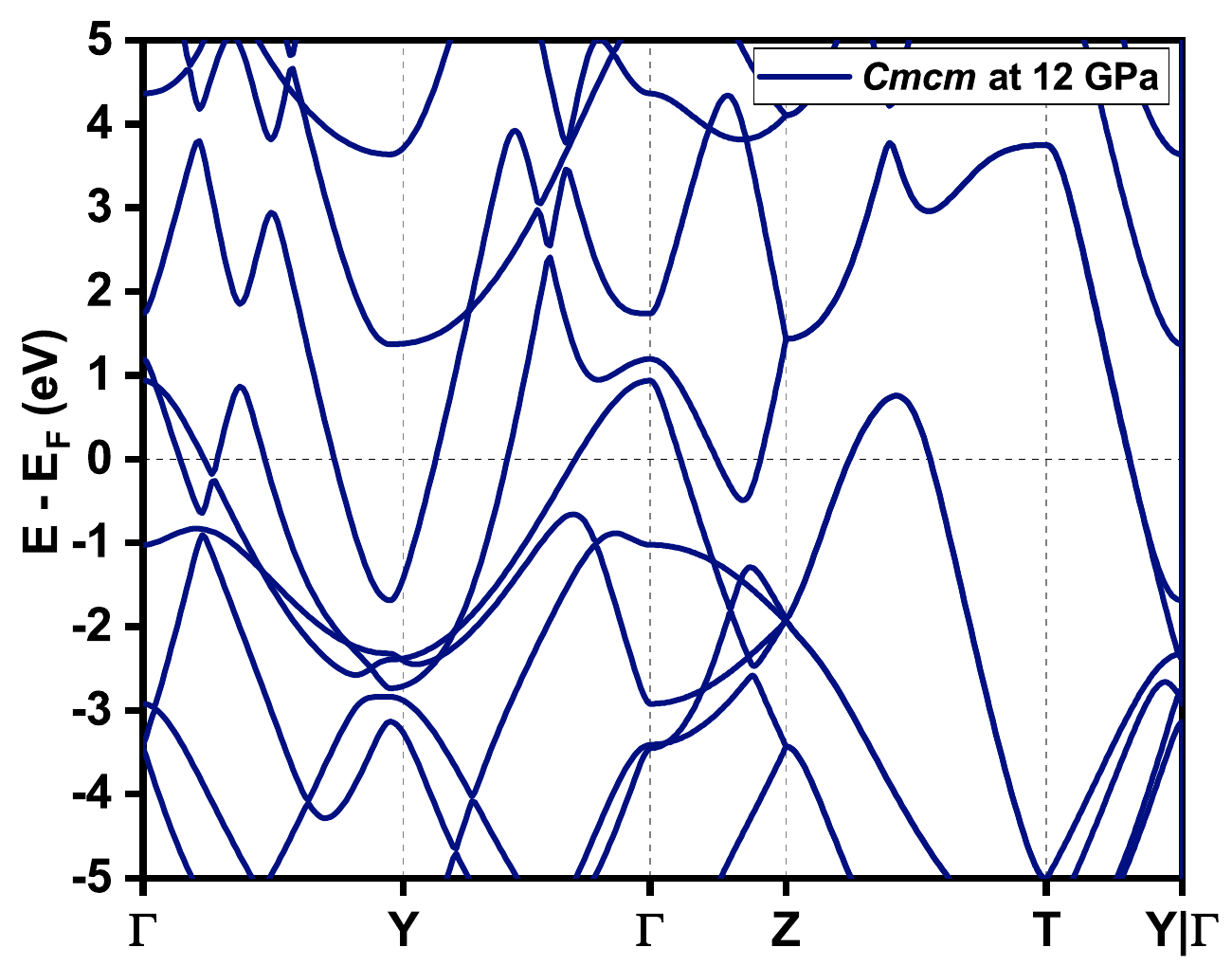}
	\end{minipage}
	\hfill
	\begin{minipage}{0.40\textwidth}
	\makebox[\linewidth][l]{\textbf{(}\textbf{d}\textbf{)}}\\[2mm]
	\includegraphics[width=\linewidth]{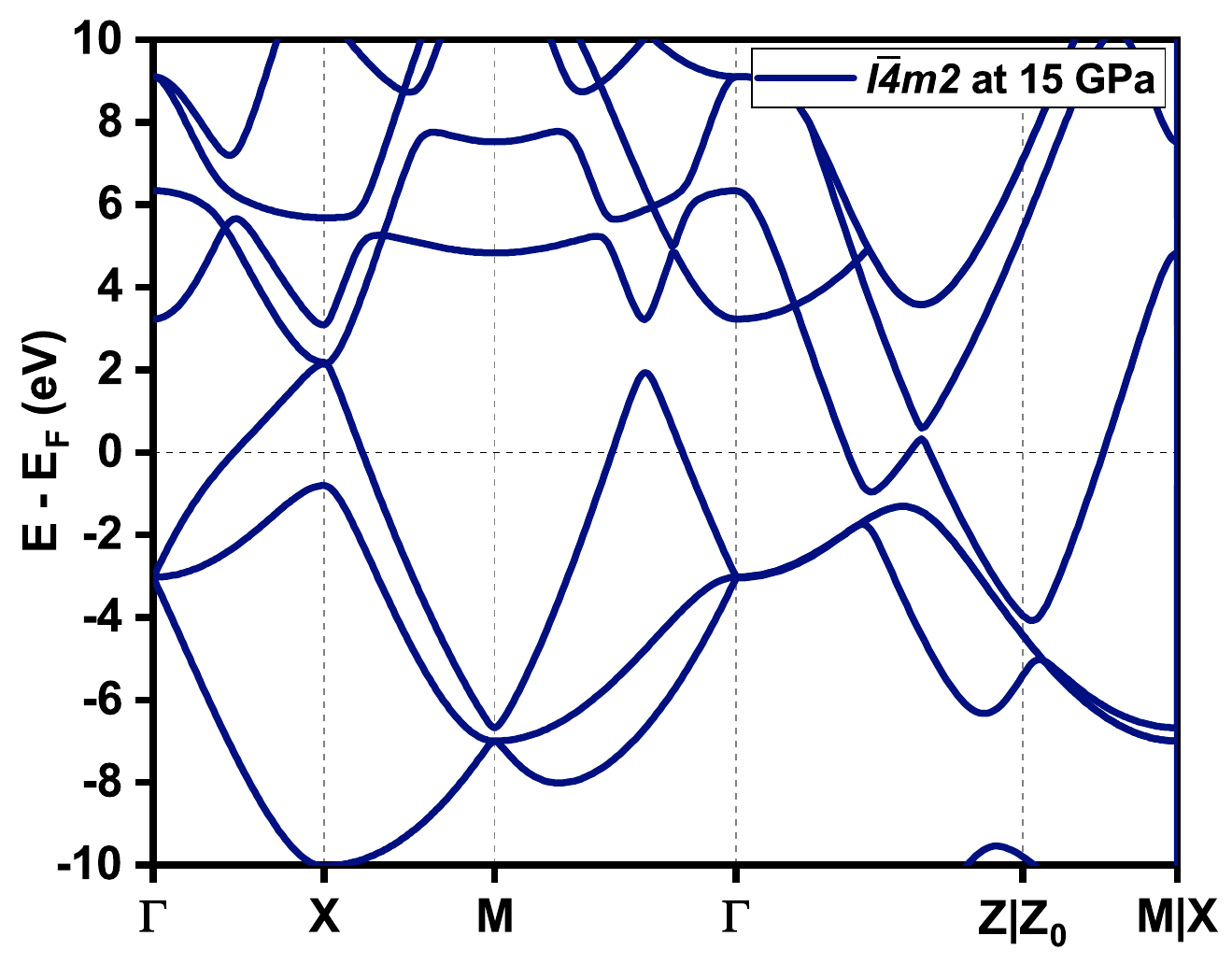}
	\end{minipage}
	
	\vfill
	\begin{minipage}{0.40\textwidth}
		\makebox[\linewidth][l]{\textbf{(e)}}\\[2mm]
		\includegraphics[width=\linewidth]{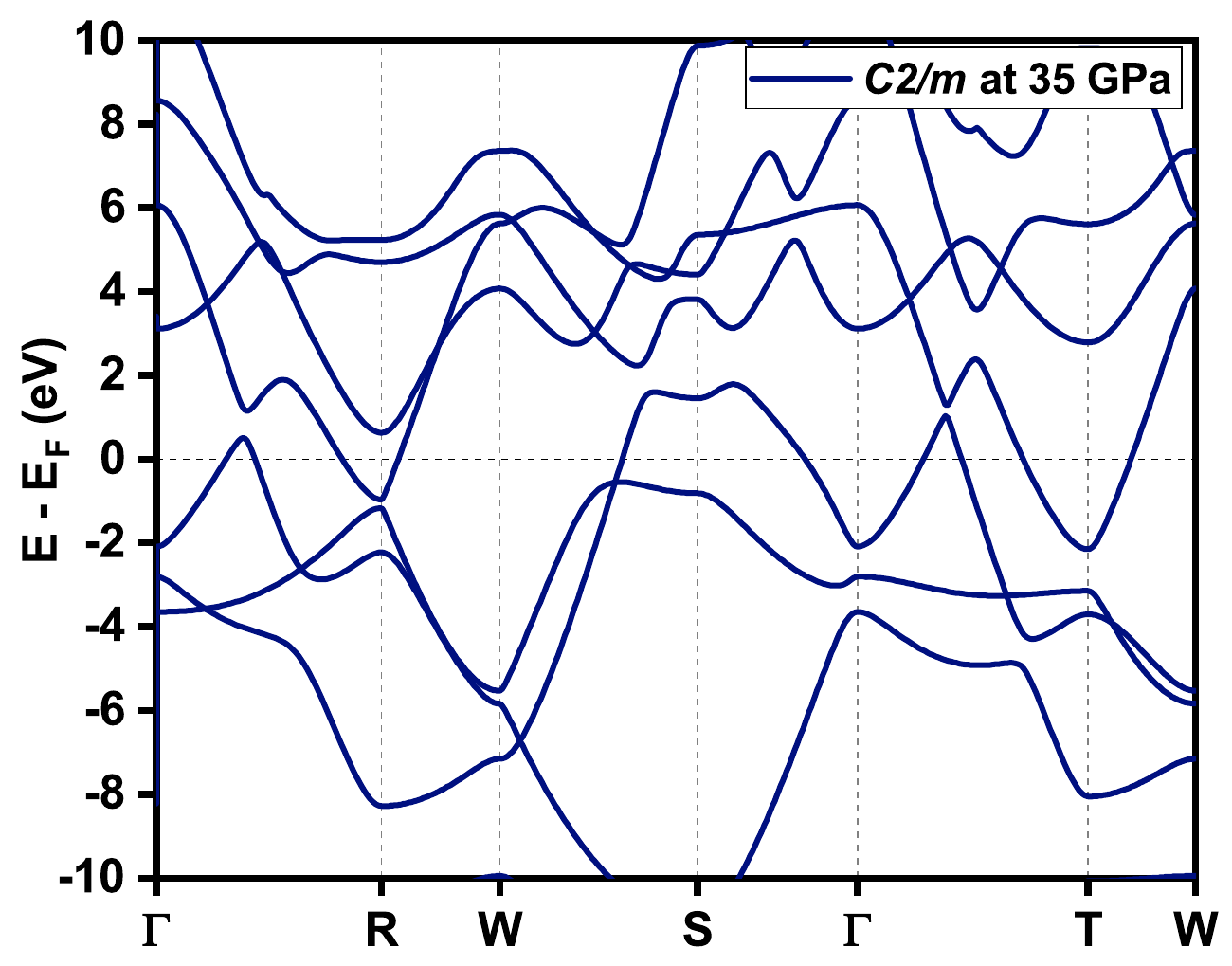}
	\end{minipage}
	\hfill
	\begin{minipage}{0.40\textwidth}
		\makebox[\linewidth][l]{\textbf{(}\textbf{f}\textbf{)}}\\[2mm]
		\includegraphics[width=\linewidth]{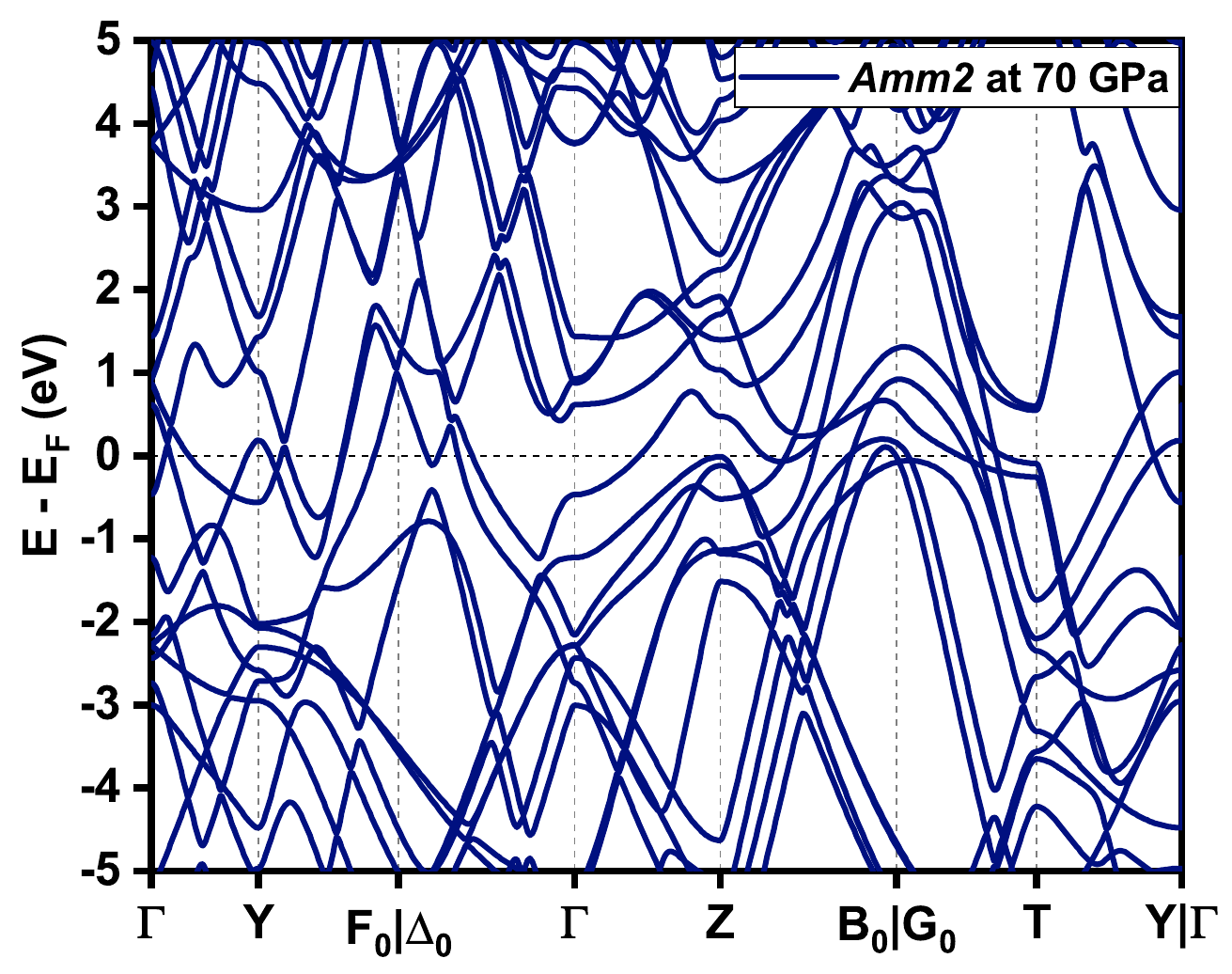}
	\end{minipage}

	\caption{Electronic band structures of considered GaAs phase at particular pressure values. (a) $F\bar{4}3m$ at 0 GPa (b) $F\bar{4}3m$ at 12 GPa (c) $Cmcm$ at 12 GPa (d) $I\bar{4}m2$ at 15 GPa (e) $C2/m$ at 35 GPa (f) $Amm2$ at 70 GPa.}
	\label{fig:Band}
\end{figure*}

\clearpage

\begin{figure}[h]
	\centering
	
	\begin{minipage}{0.60\columnwidth} 
		\raggedright 
		{\fontsize{12}{14}\selectfont} \ 
		\includegraphics[width=\linewidth]{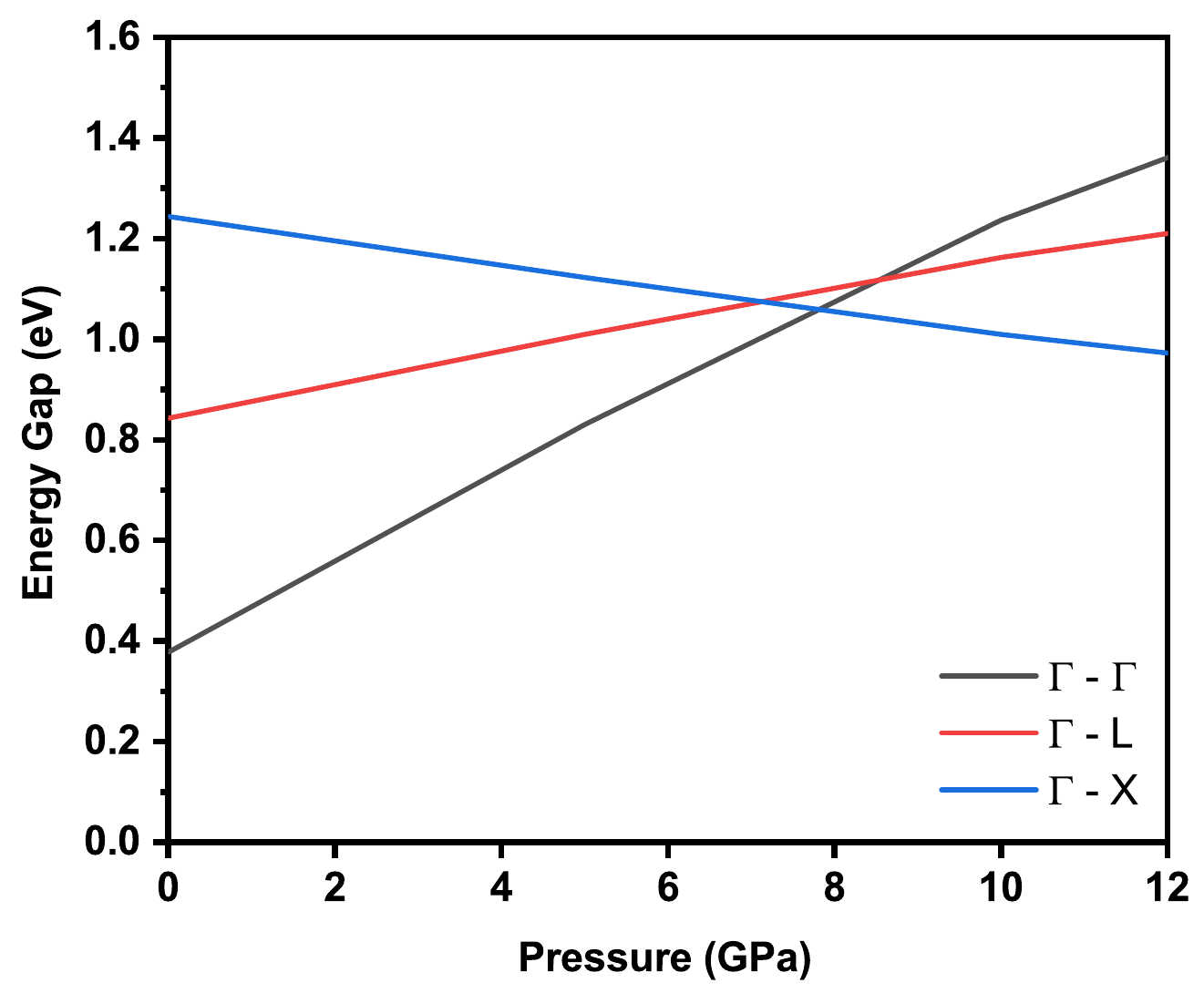} 
	\end{minipage}
	\caption{Pressure dependence of the calculated electronic band gaps of ZB GaAs for the $\Gamma$-$\Gamma$ direct transition and the $\Gamma$-L and  $\Gamma$-X indirect transitions.}
	
	\label{fig:band}
\end{figure}

\subsection{Electronic Density of States}

\begin{figure*}[h]
	\centering
	
	\begin{minipage}{0.42\textwidth}
		\makebox[\linewidth][l]{\textbf{(a)}}\\[2mm]
		\includegraphics[width=\linewidth]{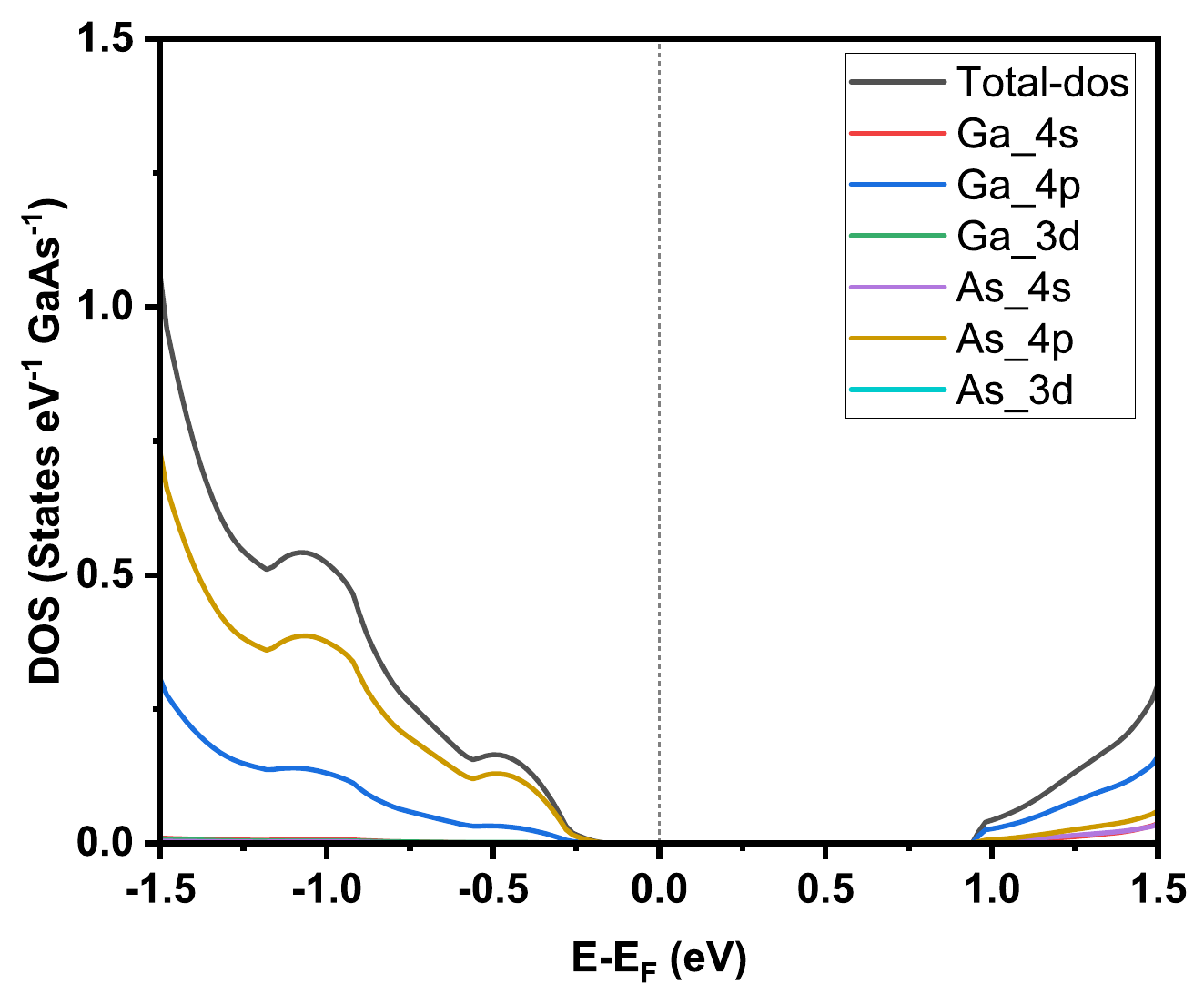}
	\end{minipage}
	\hfill
	\begin{minipage}{0.42\textwidth}
		\makebox[\linewidth][l]{\textbf{(}\textbf{b}\textbf{)}}\\[2mm]
		\includegraphics[width=\linewidth]{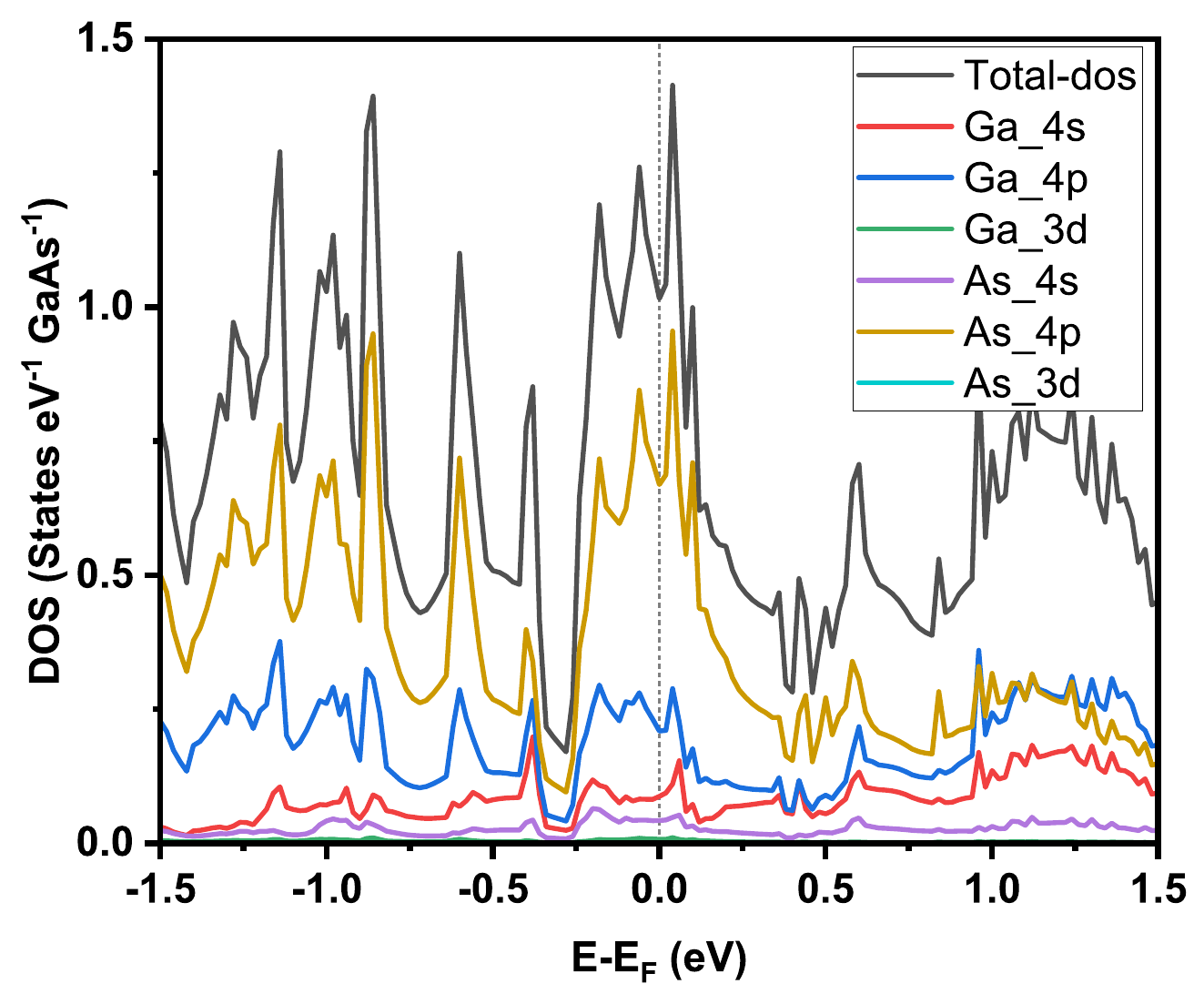}
	\end{minipage}
	\caption{Electronic density of states (DOS) of (a) $F\bar{4}3m$ and (b) $Cmcm$ at 12 GPa. In the metallic $Cmcm$, 4p states of Ga and As has major contribution at Fermi level.}
	\label{fig:DOS}
\end{figure*}

\subsection{Simulated X-ray diffraction patterns}

\begin{figure*}[h]
	\centering
	
	\begin{minipage}{0.42\textwidth}
		\makebox[\linewidth][l]{\textbf{(a)}}\\[2mm]
		\includegraphics[width=\linewidth]{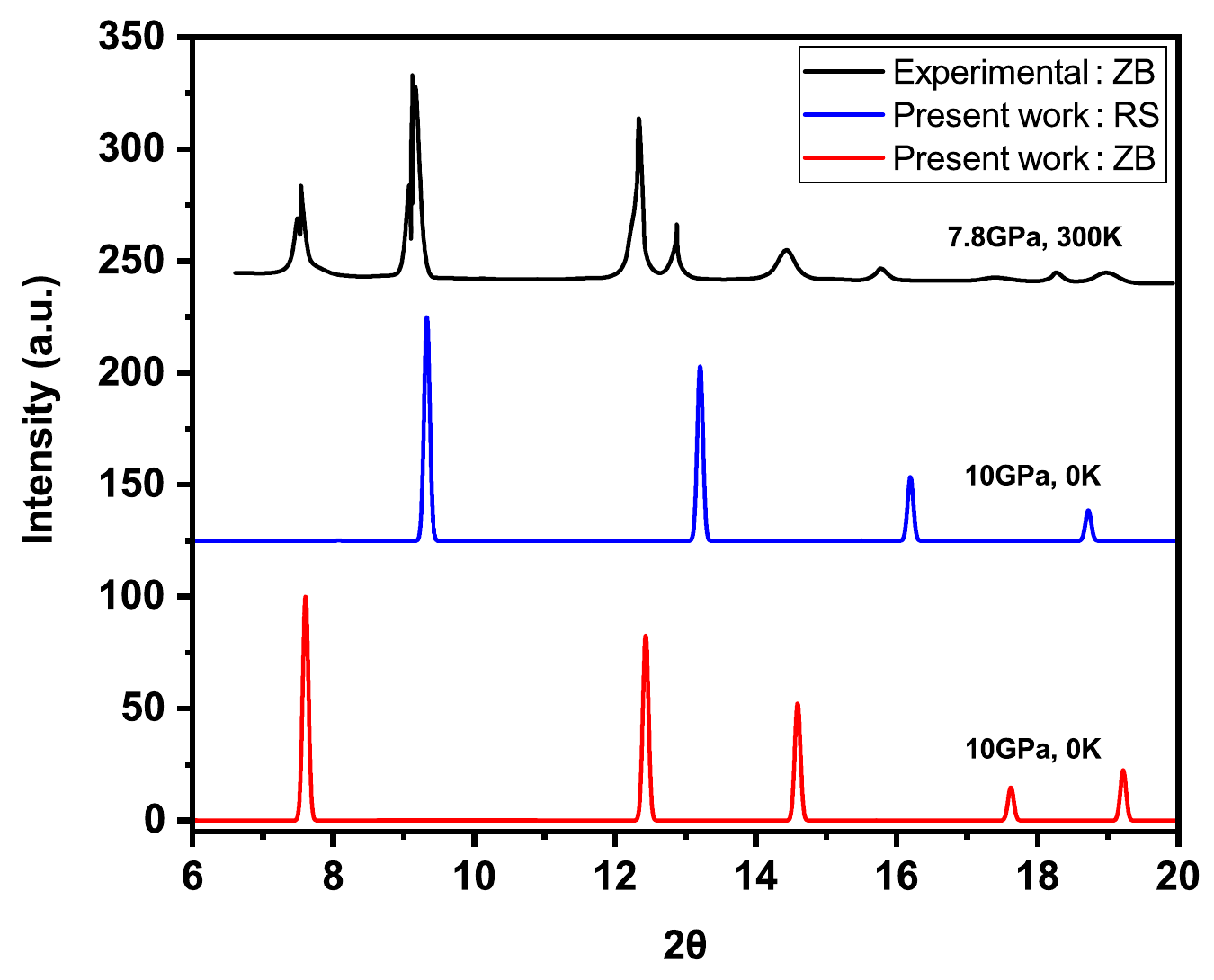}
	\end{minipage}
	\hfill
	\begin{minipage}{0.42\textwidth}
		\makebox[\linewidth][l]{\textbf{(}\textbf{b}\textbf{)}}\\[2mm]
		\includegraphics[width=\linewidth]{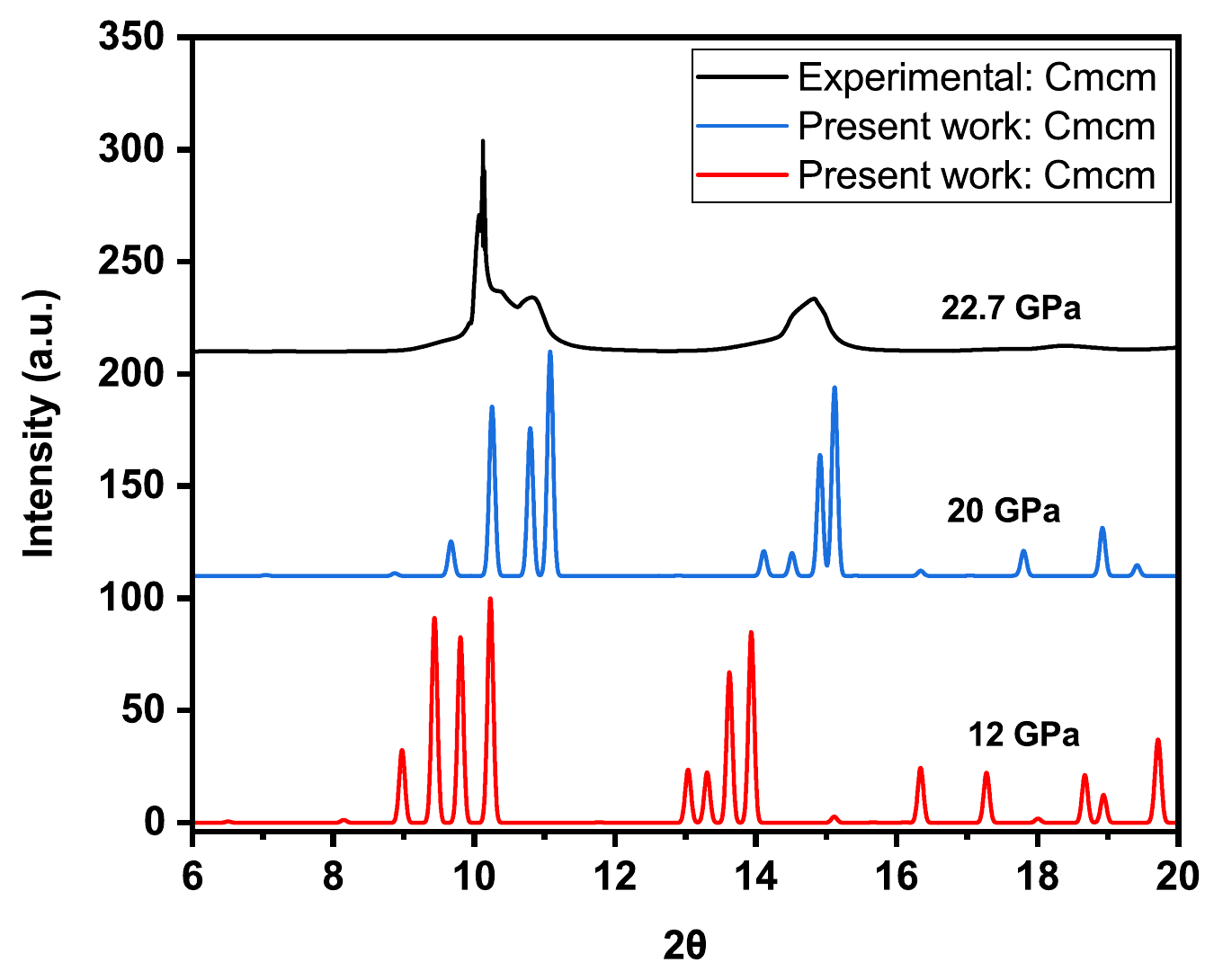}
	\end{minipage}
	\caption{(a) Comparison of the simulated XRD patterns of the ZB and rocksalt (RS) phases of GaAs, calculated at 10 GPa and 0 K, with the experimental XRD pattern of mix phases of GaAs measured at 7.8 GPa and 300 K \cite{ono2018_xray} (b) Comparison of the simulated XRD pattern of the Cmcm phase of GaAs at 12 and 20 GPa at 0 K with the experimental XRD pattern reported at 22.7 GPa \cite{McMahon_Cmcm}.}
	\label{fig:}
\end{figure*}

\begin{figure}[h]
	\centering
	
	\begin{minipage}{0.65\columnwidth} 
		\raggedright 
		{\fontsize{12}{14}\selectfont} \ 
		\includegraphics[width=\linewidth]{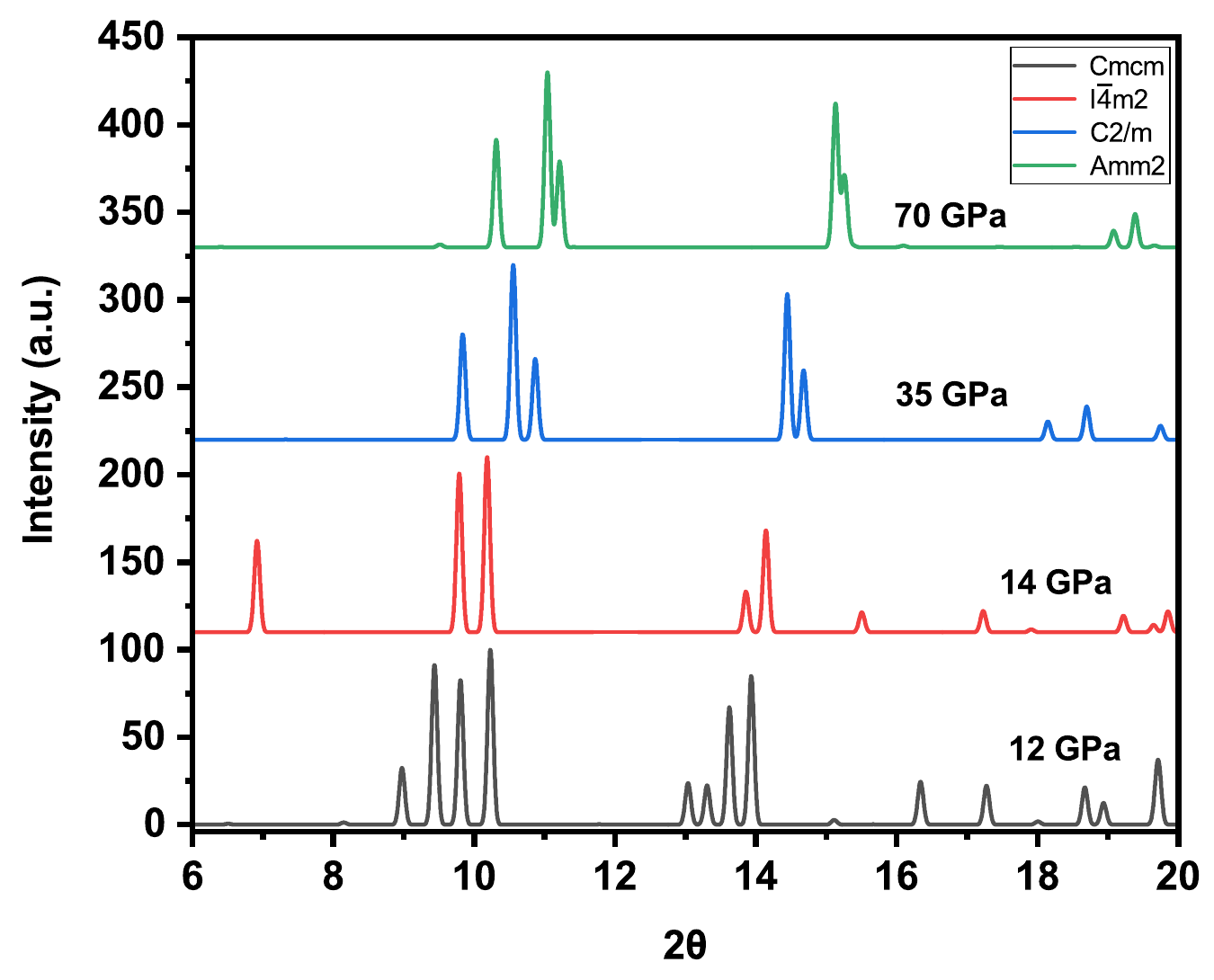} 
	\end{minipage}
	\caption{Simulated XRD patterns of the high-pressure phases $Cmcm$, $I\bar{4}m2$, $C2/m$, and $Amm2$, at pressures of 12, 14, 35, and 70 GPa, respectively. The distinct diffraction peak positions and distributions reflect the structural differences among the four phases.}
	
	\label{fig:}
\end{figure}
\clearpage
\bibliography{ref}